\documentclass[usenatbib]{mn2e}
\usepackage{graphicx}
\usepackage{natbib}
\usepackage{amsmath}
\usepackage{dsfont}
\newcommand\mC{\mbox{\bf C}}
\newcommand\vdelta{\mbox{\boldmath $\delta$}}
\newcommand\veta{\mbox{\boldmath$\eta$}}
\newcommand\vw{\mbox{\boldmath$w$}}
\newcommand\vb{\mbox{\boldmath$b$}}
\newcommand\vv{\mbox{\boldmath$v$}}
\newcommand\vk{\mbox{\boldmath$k$}}
\newcommand\vc{\mbox{\boldmath$c$}}
\newcommand\noise{{\cal N}}
\newcommand\vN{\mbox{\boldmath $\noise$}}
\def\apjl{{ApJ}}                
\def\apjs{{ApJS}}               
\def\ao{{Appl.~Opt.}}           
\def\aap{{A\&A}}                

\begin{document}

\title[Optimal mass reconstruction]{Optimal linear reconstruction of
  dark matter from halo catalogs} 
\author[Cai et al.]{Yan-Chuan Cai\thanks{E-mail: yancai@sas.upenn.edu}, Gary Bernstein and Ravi K. Sheth \\
 Department of Physics and Astronomy, University of Pennsylvania, 
 Philadelphia, PA 19104\\
 Center for Particle Cosmology, University of Pennsylvania, 
 Philadelphia, PA 19104}
\maketitle

\begin{abstract}

We derive the weight function $w(M)$ to apply to dark-matter halos 
that minimizes the stochasticity between the weighted halo distribution 
and its underlying mass density field. The optimal $w(M)$ 
depends on the range of masses being used in the estimator.  
While the standard biased-Poisson model of the halo distribution 
predicts that bias weighting is optimal, the simple fact that the mass 
is comprised of halos implies that the optimal $w(M)$ will be a mixture 
of mass-weighting and bias-weighting. In $N$-body simulations, 
the Poisson estimator is up to $15\times$ noisier than the optimal.  
Implementation of the optimal weight yields significantly lower 
stochasticity than weighting halos by their mass, bias or equal 
weighting in most circumstances.  Optimal weighting could make 
cosmological tests based on the matter power spectrum or 
cross-correlations much more powerful and/or cost-effective.
A volume-limited measurement of the mass power spectrum at
$k=0.2h/$Mpc over the entire $z<1$ universe could ideally be
done using only 6 million redshifts of halos with mass
$M>6\times10^{13}h^{-1} M_\odot$ ($1\times10^{13}$) 
at $z=0$ ($z=1$); this is $5\times$ fewer than the Poisson model 
predicts. Using halo occupancy distributions (HOD) we find that uniformly-weighted
catalogs of luminous red galaxies require $\ge 3\times$ more redshifts
than an optimally-weighted halo catalog to reconstruct the mass to the
same accuracy.  While the mean HODs of galaxies chosen to lie above a 
threshold luminosity are fortuitously very similar to the optimal $w(M)$, 
the stochasticity of the halo occupation degrades the mass estimator.  
Blue or emission-line galaxies are $\approx100\times$ less efficient at
reconstructing mass than an optimal weighting scheme.  This suggests 
an efficient observational approach of identifying and weighting halos
with a deep photo-z survey before conducting a spectroscopic survey.
The optimal $w(M)$ and mass-estimator stochasticity predicted by 
the standard halo model for $M>10^{12}h^{-1}M_\odot$ are in reasonable 
agreement with our measurements, with the
important exceptions that the halos must be assumed to be 
linearly biased samples of a ``halo field'' that is distinct from the 
mass field. Halo catalogs extending below $10^{12}h^{-1}M_\odot$ are
more stochastic than the halo model predicts, suggesting that halo
exclusion or other effects violate the assumption that halos 
sample this ``halo field'' via a Poisson process.
\end{abstract}

\begin{keywords}
  galaxies: halos  - cosmology: theory  - dark matter  - gravitational
 lensing - methods: analytical -methods: numerical
\end{keywords}

\section{Introduction}
One challenge in survey cosmology is to infer the true dark matter 
density field from observables, i.e. galaxies and galaxy clusters 
from  galaxy surveys in the UV, visible, NIR, or radio,
or galaxy clusters detected in X-ray or Sunyaev-Zeldovich (SZ) 
\citep{Sunyaev72} surveys. All these observable populations, as well 
as their host dark matter halos, are biased tracers of their underlying 
dark matter. It is well know that the bias depends on halo properties, 
e.g. mass, formation time, etc. Moreover, the tracers are
not deterministic, even on large scales, i.e. there is randomness 
in their clustering relative to that of the dark matter. 
Understanding such randomness, or stochasticity,
is crucial for precise mass reconstruction and strengthening 
constraints in cosmological parameters from observables. In this work, 
we take a step back by assuming that we have perfect ``observations''
of dark matter halos
and we aim to understand the stochasticity between halos and dark matter, 
and to develop an optimal mass estimator from halo catalogs.

The simplest assumption about the distribution of halos (or galaxies)
is that they are drawn from the mass distribution in a biased Poisson
process. This remains a common assumption, even though mass 
conservation arguments strongly suggest that this model cannot be
correct \citep{Sheth99b}.
On the other hand, the consequences of the simple truism that the
mass {\em is} the mass-weighted sum of all halos are only just beginning
to be worked out.  \citet{Abbas07} showed that the usual expressions for
halo bias \citep{MW96, ST02} -- which derive from the fact that halo
abundances are top-heavy (i.e., massive halos are overrepresented) in
denser regions---can also be derived from noting that a region which
happens to have a top-heavy mass function will also be overdense.

The two approaches provide rather different prescriptions for how
to construct an estimator of the mass field given a (subset of) the
halos.  The estimator assuming halos are Poisson sampled from the 
mass is considerably noisier than the one in which halos are mass 
weighted \citep[H10]{SHD09, Hamaus10}. In what follows, we will compare 
the differences between these two approaches, as well as provide a 
prescription for finding the optimal weight when only a subset of the 
halos are available.  

In many respects, our approach is similar to that of H10.  
However, while they concentrate on the problem of writing a weighted 
halo field as a linear function of the mass field, we will focus on 
the inverse problem -- that of writing the mass field as a weighted 
sum over the halos, and minimizing the RMS residual $E$ of the mass 
estimation.

This re-evaluation of the stochasticity in mass estimators is important
for observational programs to constrain cosmological parameters via
measurement of the mass power spectrum:  the weighting scheme that 
minimizes stochasticity may suggest a change in targeting strategies 
for spectroscopic surveys, and will reduce the resources required for 
a volume-limited measure of the mass power spectrum.
Other cosmological probes and tests of general relativity require
cross-correlation of the estimated mass field with other observables
such as gravitational lensing.  These experiments should gain even more
through the use of optimal mass reconstruction.

Section~\ref{estimators} describes previous work on how to use 
halos to estimate the mass distribution, when halos are a sampling 
of the mass field, and then contrasts these with the prescription 
which follows from assuming that the mass is mass-weighted halos.  
Section~\ref{sims} presents various tests in simulations of this 
approach.  
Section~\ref{andso} discusses implications of our findings, and 
a final section summarizes.

\section{Bias, stochasticity, and linear estimators}
\label{estimators}
Suppose we have two random variables, $m$ and $h$, both defined to
have zero mean.  We will later be interested in cases where these are
the mass and halo density fluctuations in spatial cells, or Fourier
coefficients of the fluctuation fields.  Their covariance matrix can
always be written as 
\begin{equation}
\label{pvr}
\mC = \left(\begin{array}{cc}
P & rb_{\rm var}  P \\
r b_{\rm var} P & b^2_{\rm var} P
\end{array}
\right).
\end{equation}
$r$ is the correlation coefficient between $m$ and $h$, and $b_{\rm
  var}$ is called the ``variance bias'' by \citet{DL99}.   Throughout
this paper, the variable $P$ without subscript will mean the power in
the mass distribution.  We can quantify the error in any estimator 
$\hat m$ of the mass variable via
\begin{equation}
E^2 \equiv \frac{\left\langle (m- \hat m)^2 \right\rangle}{\langle m\rangle^2}.
 \label{esq1}
\end{equation}
We will refer to $E$ as the {\em stochasticity} between the two
variables.  We can extend the definition to mean the RMS residual
error after subtraction of any estimator $\hat m$ of the mass field.
If we restrict the estimator $\hat m$ to be a linear function of $h$,
{\it i.e.\/} $\hat m = wh$ for some weight $w$, then substitution into
(\ref{esq1}) yields
\begin{equation}
E^2 = \left(1+w^2+b_{\rm var}^2-2wrb_{\rm var}\right).
\end{equation}
$E$ is minimized at $w=r/b_{\rm var}$, 
which yields
\begin{equation}
E_{\rm opt}^2 = 1-r^2 = \frac{ |\mC|}{\langle m^2\rangle \langle h^2\rangle}.
 \label{esq2}
\end{equation}

In many cases cosmological information is carried by the ratio $b$
between the two variables.  If $m$ and $h$ are drawn from a bi-variate
Gaussian distribution, then we can use the Fisher matrix formalism
from \citet{TTH} to form the covariance matrix of the parameters 
$\{P, r, b_{\rm var}\}$ in \mC.  If we draw $N_m$ pairs $(m,h)$ from 
the distribution, the uncertainties on $b_{\rm var}$ and $P$ after 
marginalization over other parameters are
\begin{eqnarray}
\frac{\sigma_P}{P} & = & \sqrt\frac{2}{N_m}\\
\frac{\sigma_{b_{\rm var}}}{b_{\rm var}} & = & \sqrt\frac{1-r^2}{N_m} 
 = \frac{E_{\rm opt}}{\sqrt{N_m}} 
 = \sqrt{\frac{E_{\rm opt}^2}{2}}\, \frac{\sigma_P}{P}.
\end{eqnarray}
[See \citet{SCS05} for the real space version of this argument.]  
Note that the same stochasticity $E$ appears in 
$\sigma_b$.\footnote{\citet{Seljak04} and \citet{Bonoli09} write 
  $\sigma_b/b=S=\sqrt{2(1-r)}$ per mode.  The difference between this 
  and the correct expression $\sqrt{1-r^2}$ is small when $r$ is 
  close to 1.}  
The appeal of cosmological tests based on ratios like $b_{\rm var}$
instead of power variables like $P$ is that the former require 
$E_{\rm opt}^2/2$ fewer mode measurements to reach the same accuracy and 
can hence be more powerful for a given survey.  We will show below that 
this factor can reach $E_{\rm opt}^2/2 < 10^{-3}$ for estimation of the 
mass distribution from halo distributions.

In this paper we will adopt a {\em covariance} definition of bias for
variable $h$ against $m$, in which the covariance matrix is written as
\begin{equation}
\label{bcov}
\mC = \left(\begin{array}{cc}
             P & b P \\
           b P & b^2P + \noise
       \end{array}\right).
\end{equation}
The ``noise'' component is $\noise\ge 0$.  Note that our $P$ is explicitly 
the mean power spectrum in realizations of the field $m$.  
We do not subtract shot noise.  

From (\ref{esq2}), the stochasticity of $h$ and $m$ is
 $E_{\rm opt}^2 = \noise/(b^2P + \noise) = (1+b^2P/\noise)^{-1}$.  
If $h$ is a Fourier coefficient of the fluctuations in a biased 
Poisson sampling of the field $m$, with mean space density $n$, then 
$\noise=1/n$.  For any field $h$, therefore, we can define an effective 
space density via
\begin{equation}
 \label{nb2eff}
 (nb^2)_{\rm eff} P = E_{\rm opt}^{-2} -1.
\end{equation}

Recently, H10 concentrated on optimization of the quantity 
\begin{equation}
\sigma_w^2 \equiv \left\langle (\delta_w - \hat \delta_w)^2 \right\rangle,
\end{equation}
where $\delta_w$ is a weighted halo map and $\hat \delta_w$ is a
linear function of the mass fluctuations.
Our approach (and motivation) is similar, although we focus on the 
inverse problem -- that of writing the mass field as a weighted sum 
over the halos, and minimizing the RMS residual $E$ of the mass 
estimation.

\subsection{Multidimensional linear estimators}
Now consider attempting to estimate the mass fluctuation $\delta_m$
from the fluctuations $\delta_i$ of a collection of tracers, {\it e.g.} 
the fluctuations in bins of halo mass.  We wish to construct an 
estimator $\hat\delta_m$ from a linearly weighted sum of the tracers: 
\begin{equation}
 \hat \delta_m \equiv \sum w_i\delta_i = \vw\cdot \vdelta.
\end{equation}
With complete generality we can define the power $P$, the covariance
bias vector ${\bf b}$ and the halo covariance matrix \mC\ via
\begin{eqnarray}
\langle \delta_m^2 \rangle & = & P, \\
 \label{covbias}
 \langle \delta_m \delta_i \rangle & = & b_iP, \\
 \langle \delta_i \delta_j \rangle & = & C_{ij}.
\end{eqnarray}
Note that we have not subtracted a shot noise contribution from the
halo variance.  We will never subtract a Poisson shot noise term from
the power spectra in this paper because it is our intention to test
the simple assumptions about the nature of shot noise.

For any choice of weight vector \vw, the stochasticity of the resulting
mass estimator is
\begin{equation}
E^2 = 1 - 2\vb^T \vw + \vw^T \mC \vw / P.
\end{equation}
The choice of \vw\ that minimizes the stochasticity is
\begin{equation}
 \label{wopt}
 \vw_{\rm opt}  =  \left(\mC/P\right)^{-1} \vb ,
\end{equation}
making
\begin{equation}
 \label{eopt}
 E^2_{\rm opt}  =  1 - \vb^T \left(\mC/P\right)^{-1} \vb.
\end{equation}
Note that this is the stochasticity-minimizing linear estimator
quite generally, independent of any assumptions about Gaussianity or
any details of the process generating the halo distributions. Linear
estimators which minimize the RMS residual of a target are known as
Wiener filters, and our form  
$\vw = \mC^{-1}\vb$ is typical for such cases.
Also it is generally true that the optimized mass estimator
 $\hat \delta_{\rm opt}$ will satisfy
 $\langle \hat \delta_{\rm opt}^2 \rangle 
 = \langle \hat \delta_{\rm opt} \delta_m \rangle$.  

The weight Eq.~(\ref{wopt}) is proportional to that in eqn.~(19)
of H10, even though their derivation assumes Gaussianity and is not
based on minimizing the stochasticity $E$.

\subsection{Principal components}
\citet{Bonoli09} investigate the stochasticity of the halo field with
respect to the matter by taking the weight function to be the first
(or higher) principal component (PC) of the halo covariance matrix \mC.  
In other words the weight \vw\ is the eigenvector of \mC\ having the
largest eigenvalue.  If the $\delta_i$ are rotated into the principal
components, then the matrix \mC\ becomes diagonal.  If principal
component $j$ has correlation coefficient $r_j$ with respect to the
matter, then it is easy to see that
\begin{equation} 
E^2_{\rm opt} = 1 - \sum_j r_j^2.
\end{equation}
Hence a drawback of PC weighting is that the stochasticity of the
first principal component (PC1) achieves the optimally low value only
if no other PC's correlate with the mass.  \citet{Bonoli09} show that 
this is a good approximation only on the largest scales.

Another issue with PC weighting is that it is not stable to re-binning
of the halo population.  For example, when the halos occupy the mass
distribution with a Poisson process, the bins must be chosen with
equal $n_i$ in order for the first PC to dominate the correlation with
the mass (as was done by H10).  The optimal weighting Eq.~(\ref{wpoi}) 
shown later in our paper is recovered only in the limit of vanishing shot noise, 
$n_iP\rightarrow\infty$.  Since PC weighting is binning-dependent and 
non-optimal, we will not focus on it.

H10 find that the optimal weight vector is very close to the weakest
principal component of the ``shot noise matrix'' $\mC - \vb P\vb^T$
when the halos are binned in equal numbers. This is intriguing since
there is no algebraic requirement for the correspondence.  

\subsection{Mass as mass-weighted halos}

\subsubsection{Mass completeness relation}
If we partition {\em all} of the mass into halos, {\it i.e.} extend
the halo catalog to zero mass, then the mass distribution {\em is} the
mass-weighted sum of the halo distributions \citep[e.g.,][]{Abbas07}.  
Hence we will obtain a perfect $E=0$ estimator of mass if we weight 
each halo bin by the fraction of the total mass it holds:
\begin{equation}
\label{w=m}
 \eta_i = \frac{n_i m_i}{\sum_i n_i m_i} = \frac{n_i m_i}{\bar\rho},
\end{equation}
where $m_i$ is the mean mass of halos in bin $i$ and $\bar\rho$ is the
overall mean density.  Since $E=0$ is clearly the optimal result, it
is mass weighting which is optimal.  If, on the other hand,
halos occupy the mass distribution via a biased Poisson process,
it is optimal to weight halos by their bias factors (we show this 
explicitly below).
Therefore, the biased Poisson model cannot be correct in the limit 
that the halo catalog includes all of the mass.

The simple truism that the mass {\em is} the mass-weighted sum of 
all halos suggests that the optimal estimator will tend toward mass
weighting as we include lower-mass halos.  \citet{Park09} note in N-body
simulations that mass-weighted halo catalogs attain lower stochasticity
than uniform weighting.  \citet{SHD09} show that weighting by mass (or
other functions of mass) produces significantly lower stochasticity
than expected from Poisson shot noise.  H10 derived that weight 
function which optimizes $\sigma_w^2$.  We derive the analogous 
optimal weight for $E$.

\subsubsection{Mass estimation with incomplete halo catalogs}
Suppose we are given a list of halo masses and positions.
Suppose that this list is complete down to some limiting mass $m_d$.  
We can assign the remainder of the mass to a ``dust bin,''  which 
contains the fraction $\eta_d = 1 - \sum \eta_i$ of mass that is 
not in the halos.  

If $\delta_d$ is the relative density fluctuation field of the mass 
in the dust bin, we can define the power of the dust field as 
\begin{equation}
   P_d = \langle \delta_d^2 \rangle
\end{equation}
and a bias vector $\vc_d$ between halos and dust by
\begin{equation}
 c_{di} = \frac{C_{di}}{P_d} = \frac{\langle \delta_d \delta_i \rangle}{P_d} .
\end{equation}
Because 
\begin{equation}
 \delta_m \equiv \eta_d \delta_d + \sum \eta_i \delta_i,
\label{msum}
\end{equation}
the bias and optimal weighting for the halo bins against the mass are:
\begin{eqnarray}
     P & = & \eta_d^2 P_d + 2 \eta_d P_d \veta^T \vc_d + \veta^T \mC \veta, \\
 P \vb  & = & \eta_d P_d \vc_d + \mC \veta, \\
\label{smoothw}
 \vw_{\rm opt} & = & \veta + (\eta_d P_d) \mC^{-1} \vc_d, \\
 \label{smoothE}
 E^2_{\rm opt} & = & \frac{ \eta_d^2}{ P/P_d} \,
                   \left( 1 - \vc_d^T \mC^{-1} \vc_d P_d \right).
\end{eqnarray}
In this formulation it is apparent that as our halo catalog extends to
lower masses and $\eta_d \rightarrow 0$, the optimal weight
 $\vw_{\rm opt}\rightarrow \veta$, {\it i.e.} mass weighting, and
 $E^2\rightarrow 0$.   
The further question of interest for finite $\eta_d$ is:  How well do
the known halo  
fluctuations $\delta_i$ predict the dust bin density $\delta_d$?

\subsection{Sampling Models}
\label{biasedPoisson}
So far the analysis has been completely general as to the generation
of the mass field and the designation of halos within it.  Now we 
examine models for the relation between halos and mass.

The most common assumption about the distribution of halos (or galaxies) 
is that they are drawn from the mass distribution in a biased Poisson 
process.  If we are examining the Fourier coefficients of the mass and 
halo distributions, the biased Poisson model can be broken down into 
three assumptions:
\begin{enumerate}
\item The halos are a linearly biased sampling of some continuous ``halo
 field'' $\delta_h$ with power $\langle \delta_h^2 \rangle = P_h$, 
  via some stochastic process that has no spatial correlations. 
  Then the covariance matrix of halo bins can be written as a rank-one
  matrix plus a diagonal ``shot noise:''
\begin{eqnarray}
\label{vvt}
\mC & = & P_h \vv \vv^T + {\rm diag}(\vN), \\
\noise_i & = & f_i / n_i.
\end{eqnarray}
Here $f_i>0$ is a ``clump size'' factor relating the noise $\noise_i$ in bin $i$ to the
space density $n_i$ of sources in the bin.  Halos in bin $i$ are drawn
with bias $v_i$ from  the halo field.  
\item The halos are placed by a Poisson process so that $f_i=1$.
\item We identify the halo field $\delta_h$ with the mass $\delta_m$
  such that $P_h=P$ and $\vv=\vb$.  In this case
 $\mC = P\vb\vb^T + {\rm diag}(1/n_i)$.
\end{enumerate}

Assumption (iii) of the biased-Poisson model has been noted to be
inconsistent with the assumption that the halo catalog can be extended
to comprise all of the mass.  We will therefore consider models in
which assumption (i) holds without (iii), such that the halos sample a
field that is distinct from the mass distribution.  We will take care
therefore to distinguish \vv, the bias of the halos with respect to
the halo field $\delta_h$, from \vb, which we always define via the
covariance with mass as per (\ref{bcov}).

The assumptions that halos are linearly biased, and that the halo 
generating process has no spatial correlations are idealizations:  
in fact, halos in simulations do not overlap (almost by definition), 
and their bias is non-linear.  We will return to the limits of these 
assumptions later.

When \mC\
takes the form (\ref{vvt}), two things are of note: first, this
description is stable under re-binning of the halos in the limit of
narrow bins.  More specifically, if $v_i$ and $f_i$ are slowly-varying
functions of the mass $m_i$ of halos in bin $i$, then the
$v_i$ and $f_i$ do not change if two adjacent bins are merged.  In
other words we can write functions $v(m)$ and $f(m)$, and all of our
linear-algebra formulations can be carried over into integrals over
halo mass $m$.  The second useful fact about (\ref{vvt}) is that it 
can be inverted analytically using the Sherman-Morrison formula:
\begin{eqnarray}
\label{vvtinv}
P_h\left(\mC^{-1}\right)_{ij} & = & 
  x_i \delta_{ij} - \frac{ x_i v_i x_j v_j}{1 + \sum x_i v_i^2} \\
  x_i & \equiv & P_h/\noise_i = n_i P_h/f_i.
\end{eqnarray}

When all three conditions of the biased-Poisson model are met, 
the optimal weight function and
stochasticity are found simply from (\ref{wopt}) and (\ref{vvtinv}):
\begin{eqnarray}
\label{wpoi}
 \frac{w_{{\rm opt},i}}{n_i} & = & b_i \frac{P}{1+\sum n_i b_i^2 P}, \\
 \label{epoi}
 E^2_{\rm opt} & = & \left(1 + \sum n_i b_i^2 P \right)^{-1}, \\
\label{nbpoi}
(nb^2)_{\rm eff} & = & \sum n_i b_i^2.
\end{eqnarray}
This recovers the result from \citet{Percival04} that the optimal
linear mass estimator in a Poisson model weights each halo by its bias
$b_i$ (times a mass-independent factor), and the stochasticity of the
estimator is determined by $\sum n_i b_i^2 P$.
Conveniently the weights scale with the bias, independent of the range
of halo masses included in the estimator.  This property does 
{\em not} hold for more general forms of \mC\ and \vb, 
{\it e.g.} it fails when $\vv \ne \vb$ and condition (iii) is violated.

In this paper we will {\em not} assume that the halos occupy the
matter distribution via a biased Poisson process.  We will examine the
\mC\ matrix for halos in numerical simulations, examine what if any
of the three biased-Poisson conditions actually holds, and then use
the general formulae (\ref{wopt}) and (\ref{eopt}) to find how the
optimal stochasticity differs from the Poisson predictions.

\subsubsection{General sampling model}
\label{samplingmodel}
We now examine the case where all halos, and the dust, are indeed 
placed by a local process that is biased relative to some halo field
$\delta_h$, so assumption (i) holds but (ii) and (iii) are not
assumed.
This model illustrates the difference between the halo field that 
is {\em sampled} and the mass field that the halos {\em comprise}.  
The two fields cannot be equivalent, even in the linear regime.
In \S\ref{testsampling}, we examine whether halo covariance matrices measured 
in simulations are in fact consistent with this sampling model.  

If all the halos and dust are placed in this halo field by
independent processes, and we define a weighted field
\begin{equation}
\delta_v \equiv \frac{\sum_i n_i v_i \delta_i}{\sum_i n_i v_i},
\end{equation}
then the covariances between the dust, the halo bins, and $\delta_v$ 
are
\begin{eqnarray}
\label{vvNmodel}
        \mC & = & \vv \vv^T P_h + {\rm diag}(f_i/n_i) \\
 P_d = C_{dd} & = & v_d^2 P_h + \noise_d \\
       C_{di} & = & v_d v_i P_h \\
C_{dv}  & = & v_d \frac{\sum_i n_i v_i^2}{\sum_i n_iv_i} P_h\\
 C_{vv} & = & \left(\frac{\sum_i n_i v_i^2}{\sum_i n_iv_i}\right)^2 P_h 
            + \frac{\sum_i n_i v_i^2 f_i}{(\sum_i n_i v_i)^2}
\end{eqnarray}

We have freedom in setting the normalization of $v_d$ and \vv\ which we 
do by requiring 
\begin{equation}
 \label{normV}
 \eta_d v_d + \veta \cdot \vv = 1.
\end{equation}
In this case the mass, which is the sum of halo and dust densities,
has power
\begin{eqnarray}
\label{PmPh}
P= C_{mm} & = & P_h + \eta_d^2\noise_d + \sum \eta_i^2 f_i / n_i \\
       & = & P_h + \eta_d^2\noise_d 
               + (1-\eta_d)^2 \langle fm^2\rangle / \langle m\rangle^2 \\
       & \equiv & P_h + \noise_m.
\end{eqnarray}
The angle brackets denote number-weighted averages over the halo
population, and the final expression defines $\noise_m$.  
We see that when the mass field is a sampled realization of the halo 
field, then $P\equiv C_{mm}$ is larger than $P_h$ by terms representing 
the sampling shot noise.  

The bias \vb\ of the halo mass bins relative to the mass distribution 
will not in general equal the bias \vv\ with respect to the halo
field.  Since $b_i P = \langle \delta_i \delta_m\rangle,$ we can expand
$\delta_m$ using (\ref{msum}), and use the $\mC$ elements above, to
derive 
\begin{equation}
 \label{vbsample}
 \vb = \frac{\vv + {\rm diag}(m_i f_i/P_h\bar\rho)}{P/P_h}.
\end{equation}
Formally, only $v_i \propto f_i m_i$ will yield $\vb \propto \vv$.  
In general, $b_i>v_i$ at sufficiently large masses, and $b_i<v_i$ at 
lower masses.  Equality is at $m_i = b_i \noise_m\bar\rho$, which 
occurs for $z=0$ at $\sim 10^{14} h^{-1}M_\odot$.  In a $\Lambda$CDM
model, the $b=v$ crossover will occur for halos with $b\approx1.6$ for
a wide range of redshifts. 
We can also solve for the weight vector of the optimal linear mass
estimator:
\begin{equation}
\label{woptsample}
\frac{w_{{\rm opt},i}}{n_i} = \frac{m_i}{\bar\rho} 
          + \eta_d v_d\,(v_i/f_i)\, P_h\,E^2_{\rm pois},
\end{equation}
where we have set 
$E^2_{\rm pois} \equiv (1 + \sum_j n_j v_j^2 P_h/f_j)^{-1}$;
the sum over $j$ is over the halo bins in the catalog.  $E^2_{\rm
  pois}$ describes the fidelity with which the halos can estimate the
{\em halo} field, as opposed to the {\em mass} field.
Notice that the first term is the mass weighting 
$(w\propto m)$, and the second correction term has the same form 
as that for the standard Poisson model $(w\propto v)$, but its 
importance depends on the mass fraction in dust, and how it clusters.  
Thus, equation~(\ref{woptsample}) is a weighted sum of 
equations~(\ref{w=m}) and~(\ref{wpoi}). 
As the halo catalog comprises more of the total mass,
$\eta_d\rightarrow 0$, and mass weighting becomes optimal.  

If we define $\delta_{\rm opt} \equiv \sum_i w_{{\rm opt},i}\delta_i$, then 
the optimized stochasticity is 
\begin{eqnarray}
 \label{E2opt}
 E^2_{\rm opt} &=& 1 - \frac{\langle\delta_{\rm opt}\delta_m\rangle^2}
                 {\langle\delta_{\rm opt}^2\rangle\langle\delta_m^2\rangle} 
              =  1 - \frac{\langle\delta_{\rm opt}\delta_m\rangle}
                          {\langle\delta_m^2\rangle} \\
     &=& \frac{\eta_d^2C_{dd}}{C_{mm}} \,
          \left(1 - \frac{C_{dv}^2}{C_{dd}C_{vv}}\right)\\
      &=& \frac{\eta_d^2}{P_m} \left( \noise_d + v_d^2 P_h\,E^2_{\rm pois}\right).
\label{E2sample}
\end{eqnarray}
We have written the second equality explicitly to show that $E_{\rm opt}$ 
makes physical sense:  since $\eta_d^2 C_{dd}/C_{mm}$ is the fraction of 
the total $C_{mm}$ that is in dust, the stochasticity is the yet
smaller portion
of this dust power that cannot be recovered via the correlation between 
the dust and the bias weighted halos.  

The final expression shows that $E_{\rm opt}\to 0$ as the mass fraction 
in dust $\eta_d\to 0$, as it should.  
When $P_h\to 0$, then $E_{\rm opt}^2\to \eta_d^2\noise_d/\noise_m$:   
in this limit, the stochasticity is determined by the fraction of the 
noise term $\noise_m$ which is contributed by the dust.  
The opposite limit is when $P_h\gg \noise_m$, where $P\approx P_h$ 
and $E_{\rm opt}\to \eta_d v_d E_{\rm pois}$.  Since $\eta_d v_d < 1$, this 
is why the optimal stochasticity can be substantially smaller than 
in the Poisson model.  

The optimized weights and relations between $b$ and $v$ are similar to
what H10 found in their analysis of $\sigma_w$.  
E.g., our equation~(\ref{vbsample}) reduces to their equation~(41) 
upon taking $f_i\rightarrow 1$, $P_h\rightarrow P_{\rm lin}$, and 
$v_i\rightarrow b_i$.  
In the discussion following their Eq.~(36), H10 note that their optimal 
weight is a linear combination of mass and bias weighting, a point they 
make again with their Eq.~(49). But our expression for the relative 
contributions of these two weights is substantially more transparent 
than theirs. For instance, our formulation shows that this factor is, 
in fact, the one associated with the usual Poisson-sampling model, times 
a factor which accounts for the dust -- this is not obvious from their expressions.

\subsection{The halo model description}
\label{halomodel}
The halo model is a specific case of the sampling model in the 
previous section.  The halo model is particularly well-suited to 
describing the effect of weighting halos \citep{Sheth05}; it 
predicts not only \mC, but also the dust-bin quantities like 
$\noise_d$ and $P_{d}$ which are not observable and were left 
unspecified in the previous section.  Hence the halo model allows 
an estimate of the stochasticity $E_w$ associated with any weight 
function $w$ applied to the halos, so it can be used to estimate 
$E_{\rm opt}$.
(In the context of the optimal weight discussed earlier, it provides 
a prescription for the effect of bias weighting halos.)  

In what follows, we will explicitly set $f=1$; comparison of the 
predictions of this calculation with the measurements in simulations 
provides a measure of the accuracy of this assumption.  
In particular, if we define 
\begin{eqnarray}
 \label{nw}
 n_w &=& \int_{m_d}^\infty dm\, \frac{dn}{dm}\, w(m) ,\\
 \noise_w &=& \int_{m_d}^\infty dm\, \frac{dn}{dm}\, \frac{w^2(m)}{n_w^2}, \\
\noise_\times &=& \int_{m_d}^\infty dm\, \frac{dn}{dm}\, 
                  \frac{m\,u(k|m)}{\bar\rho}\, \frac{w(m)}{n_w}, \\
 \noise_m &=& \int_0^\infty dm\, \frac{dn}{dm}\, 
              \frac{m^2\,|u(k|m)|^2}{\bar\rho^2}
\label{noisem}
\end{eqnarray}
then
\begin{eqnarray}
 \label{pk}
 C_{ww} &=& v_w^2\, P_h(k) + \noise_w ,\\
 C_{wm} &=& v_w\, P_h(k) + \noise_\times ,
\end{eqnarray}
where 
\begin{equation}
 \label{bw}
 v_w = \int_{m_d}^\infty dm\, \frac{dn}{dm}\, \frac{w(m)}{n_w}\, v(m).
\end{equation}
Here $v(m)$ is the bias with respect to $P_h(k)$; it is related to 
the bias $b(m)$ with respect to the mass field by equation~(\ref{vbsample}).  
The factor $u(k|m)$ in the expressions above represents the fact 
that halo catalog is treated as if all mass is concentrated at the 
center of mass, but real halo mass is smeared in a density profile:  $u$ is the 
Fourier transform of the density profile, normalized so that $u\to1$ 
as $k\to 0$.  

Strictly speaking, this writing of the halo model is not quite 
correct, because halos of a given mass may have a range of density 
profiles,  i.e. $u$ is not the same for all halos, and stochasticity 
in $u$ will contribute to $E$.  If we use the mean $u$ for a given $k$ 
in the expressions above, and define 
$\sigma_u^2(k|m) \equiv \langle |u^2| \rangle - \langle u\rangle^2$, 
then then the scatter in profile shapes will contribute an 
additional term 
\begin{equation}
\label{structurenoise}
\noise_m \rightarrow \noise_m + \int dm \frac{dn}{dm} \sigma_u^2(k|m)
\left(\frac{m}{\bar \rho}\right)^2.
\end{equation}
We expect this additional term to be unimportant at the scales 
$k<0.1h\,{\rm Mpc}^{-1}$ that are of most interest for cosmology 
\cite{ShethJain03}.  Section~\ref{depart} shows that this is indeed 
the case.  However, this term grows as $k^4$, so it could dominate 
the stochasticity at higher $k$.  

The results of the previous section suggest that an optimal linear 
reconstruction of the mass can be obtained if we use 
equation~(\ref{woptsample}) for the weight function.  
If we set $f_i=1$, then 
\begin{equation}
 \label{halomodelw}
w_{\rm opt}(m) = \frac{m\,u(k|m)}{\bar\rho} 
          + F_v\, \frac{v(m)\, P_h(k)}{1 + (nv^2)_h P_h(k)},
\end{equation}
where 
\begin{eqnarray}
F_v &=& 1 - 
        \int_{m_d}^\infty dm\,\frac{dn}{dm}\frac{m\,u(k|m)}{\bar\rho}\,v(m) \\
 (nv^2)_h &=& \int_{m_d}^\infty dm\,\frac{dn}{dm}\,v^2(m).
\end{eqnarray}
This makes the stochasticity 
\begin{equation}
\label{halomodelE}
E_{w_{\rm opt}}^2 = 1 - \frac{C_{wm}^2}{C_{mm}C_{ww}} = 1 - \frac{n_w\,C_{wm}}{C_{mm}} 
\end{equation}
equal the expression given in equation~(\ref{E2opt}).  
(Note that, for these choices of $w$ and $v$, $n_w^2\,C_{ww} = n_w C_{wm}$, 
and $E_{w_{\rm opt}}^2$ is independent of $u$.)

Thus, we can use the halo model to estimate the stochasticity $E_w^2$ 
associated with the weight $w$ of equation~(\ref{halomodelw}) as follows.  
We can measure the mass power spectrum $P$ and the covariance bias 
$b(m)$ in the simulations, and then use equation~(\ref{vbsample}) 
to infer $v(m)P_h$.  We then use the halo model to estimate $\noise_m$, 
which we subtract from the measured $P$ to get $P_h$, and hence $v(m)$.  
These can then be used to estimate $F_\nu$ and $(nv^2)_h$ by summing over 
the halos.  The halo model can also be used to estimate $\noise_m$; 
by directly measuring the component of this that comes from the halos 
in the sample, one can determine $\noise_d$ and so estimate $E_w$.  
In the following section, we will compare this Poisson sampled and 
mass-weighted halo model for $E_w$ with the optimal stochasticity in 
simulations.

\subsubsection{Halo exclusion and other subtleties}
We could have made heavier use of the halo model as follows.  
The usual implementation \citep{Sheth05} replaces 
$v(m)\,P_h \to b_{\rm pbs}(m)\,P_{m}$, where 
$b_{\rm pbs}(m)$ is the peak background split bias \citep[][e.g.]{Bardeen86, Cole89, MW96,Sheth99},
and $P_m$ is the power spectrum of the mass, usually approximated 
by $P_{\rm lin}$ at small $k$.  
H10 make this same assumption in their halo model of $\sigma_w$.  
We will show later that $b_{\rm pbs}(m)$ appears to be closer to 
$v(m) = (C_{hm}-\noise_\times)/(C_{mm}-\noise_m)$ than it is to 
$b(m) = C_{hm}/C_{mm}$.
However, note that our discussion indicates that $P_h$ is {\em not} 
to be identified with the mass power spectrum and $v(m)$ is not the 
same as the linear bias factor between the halo and mass fields.  
This is one reason why our construction of the halo model above was slightly 
different from standard. In particular, we did not begin from the mass field 
$\delta_m$, and immediately set $P_h = P_m$, as is usually done. Rather, we 
framed our discussion in terms of the field $\delta_h$, and weighted samplings 
of it. Because we assumed that halos were linearly biased Poisson samplings of 
this field, we explicitly ignored the fact that, in reality halos are spatially 
exclusive, and the sampling function is not just a linear function of the mass. 
The exclusion property means that the assumption that the halos are obtained from 
independent sampling processes for every mass bin cannot be correct. Indeed, in 
the sampling algorithm described in \citep{Sheth99b}, this lack of independence 
appears explicitly -- and it also contributes to the non-linearity of the bias 
relation (see their equation 17). [For more recent discussion of the effects of 
halo exclusion and non-linear bias on $P_m$, and another way of seeing why exclusion 
alone can produce effects which appear as scale dependent non-linear bias, see 
\citet{Smith07}.]  As we shall see, our neglect of these effects sets a limit to 
the accuracy of our approach (e.g., the contribution to the stochasticity which 
comes from bias-weighting the halos may not be optimal).

\subsubsection{Halo-mass-dependent selection function}
The results above assume that all halos above a sharp threshold in
mass are observed.  If the threshold is not sharp, but is a
function of mass, $0\le p(m)\le 1$, then it is straightforward to
verify that equation~(\ref{halomodelw}) remains the optimal weight,
provided that, in the previous expressions for $C_{ww}$, $C_{wm}$,
$F_v$ and $(nv^2)_h$ (but not $\noise_m$, of course), all occurrences 
of $(dn/dm)$ are replaced by $(dn/dm)\,p(m)$.  As a result, the 
sub-sampling decreases $n_wC_{wm}$; since $C_{mm}$ does not change (of 
course), the sub-sampling degrades (i.e., increases) $E$.

\begin{table*}
\centering
\caption{Basic parameters used in the Millennium and NYU simulations. 
 $\epsilon$ is the Plummer-equivalent comoving softening length of 
 the gravitational force; $N_{\rm re}$ is the number of realizations for 
 each simulation; $z_{\rm start}$ is the starting redshift for the simulation.}
\bigskip
\begin{tabular}{lcccccccccccc}
\hline
Name       &$N_p$   &$M_p$ & $L_{box}$ & $\epsilon$& $\Omega_m$ & $\Omega_{\Lambda}$ &  $\Omega_b$  & $\sigma_8$ & $n_s$ &  $H_0$ & $z_{start}$ & $N_{re}$ \\ 
\hline

Millennium &$2160^3$&$8.6\times10^8~h^{-1}M_{\odot}$& 500~$h^{-1}$Mpc & 5$h^{-1}$kpc &0.25 & 0.75 & 0.045 & 0.9 & 1 & 73  & 127 & 1 \\
NYU        &$640^3$ &$6\times10^{11}~h^{-1}M_{\odot}$ & 1280~$h^{-1}$Mpc &20$h^{-1}$kpc &0.27  & 0.73 & 0.046&0.9 & 1 & 72 & 50 & 49\\
\hline
\end{tabular}
\label{tab1}
\end{table*}

\section{Optimal weights and mass estimators from simulations}\label{sims}
\subsection{Simulations}

In this section we use N-body simulations to find optimal 
mass-estimation weights for halos binned by mass (rather than, 
e.g., angular momentum, axis ratio, formation time or concentration).  
We show the stochasticity that results from applying the optimal 
weights and compare to common sub-optimal choices.

For purposes of exploration, we wish to have a wide range of halo
masses, while still having good statistics for large halos and
large-scale modes.  For the first purpose, we use the Millennium
simulation \citep{Springel05b}, which resolves halos down to 
$\sim 10^{10} h^{-1}M_{\odot}$.
For the second, we use the suite of 49 cosmological dark matter
``NYU'' simulations described in \citet{Manera10}, which have a
total volume $800\times$ larger than the Millennium, but the minimum 
resolved halo mass is $1000\times$ larger.
The simulations assume very similar $\Lambda$CDM cosmological models
and were carried on using the same {\sc GADGET-2} code \citep{Springel05a}.
Table~\ref{tab1} provides details of the basic simulation parameters.

The initial power spectrum was generated by {\sc CMBFAST} \citep{Seljak96}.  
The initial density field of the Millennium simulation was realized by 
perturbing a homogeneous, glass-like particle distribution with a 
Gaussian random realization of the initial power spectrum \citep{white96}, 
while the NYU simulations use an algorithm motivated by Second Order 
Lagrangian Perturbation Theory \citep{Scoccimarro98}.
Dark matter halos with at least 20 particles are identified in both 
simulations using the friends-of-friends (FOF) group finder
with a linking length of $0.2$ times the mean particle separation 
\citep{Davis85}. The lowest mass halo of the two 
simulations that we use are $1.7\times10^{10}h^{-1}M_{\odot}$ and 
$1.0\times10^{13}h^{-1}M_{\odot}$ respectively.   
The Millennium simulation has $\approx 1.8\times10^7$ halos at
at $z=0$, $z=0.5$ and $z=1$ while the NYU simulations have 
$\sim 4.3\times10^7$, $3.5\times10^7$ and $2.5\times10^7$ halos in 
each of these redshift slices.
For a more detailed analysis of the halo mass function and bias factors
in these simulations, see \citet{Springel05b} and \citet{Manera10}.

\begin{figure*}
\begin{center} 
\resizebox{\hsize}{!}{
\includegraphics[angle=0]{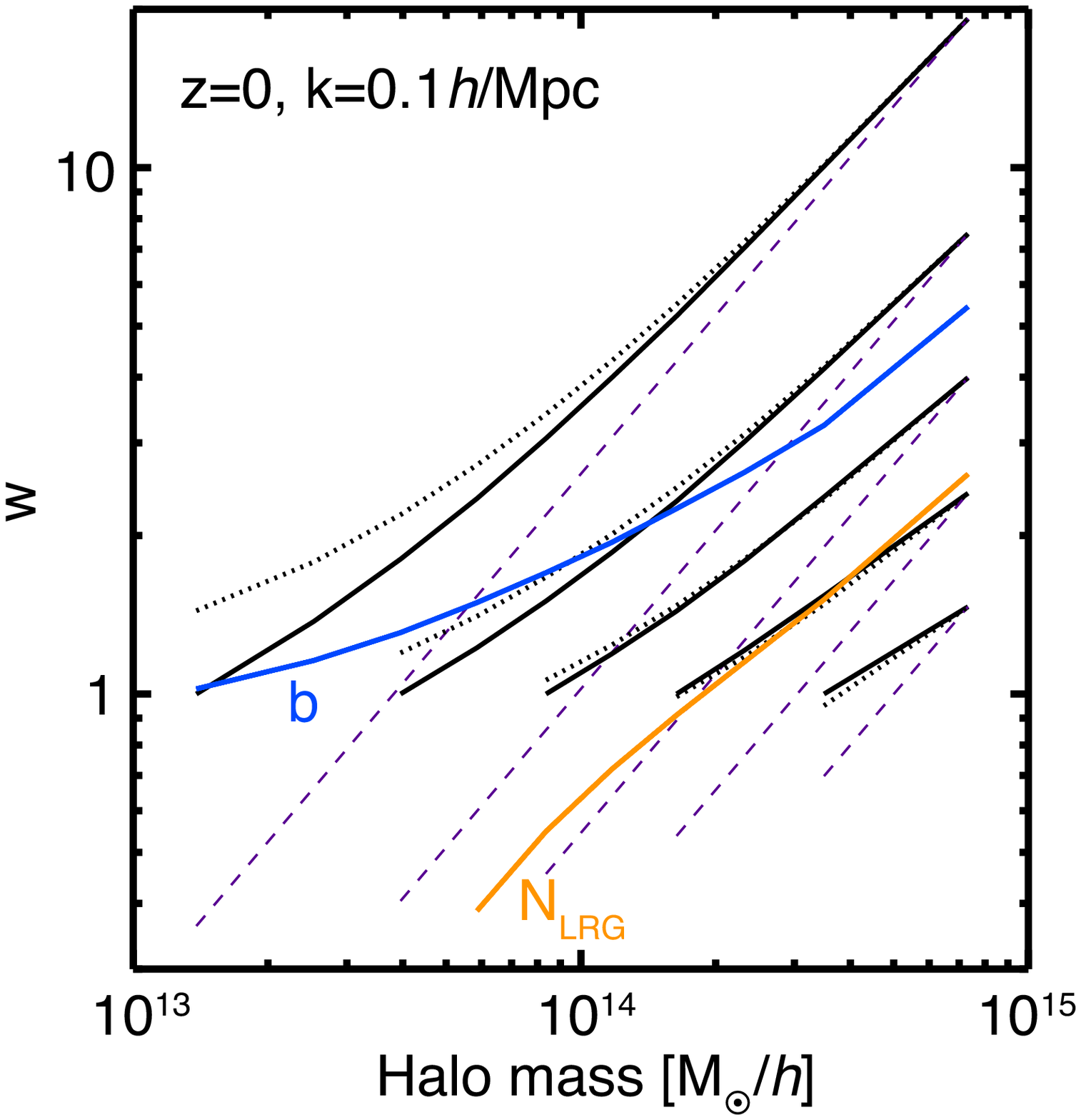}
\includegraphics[angle=0]{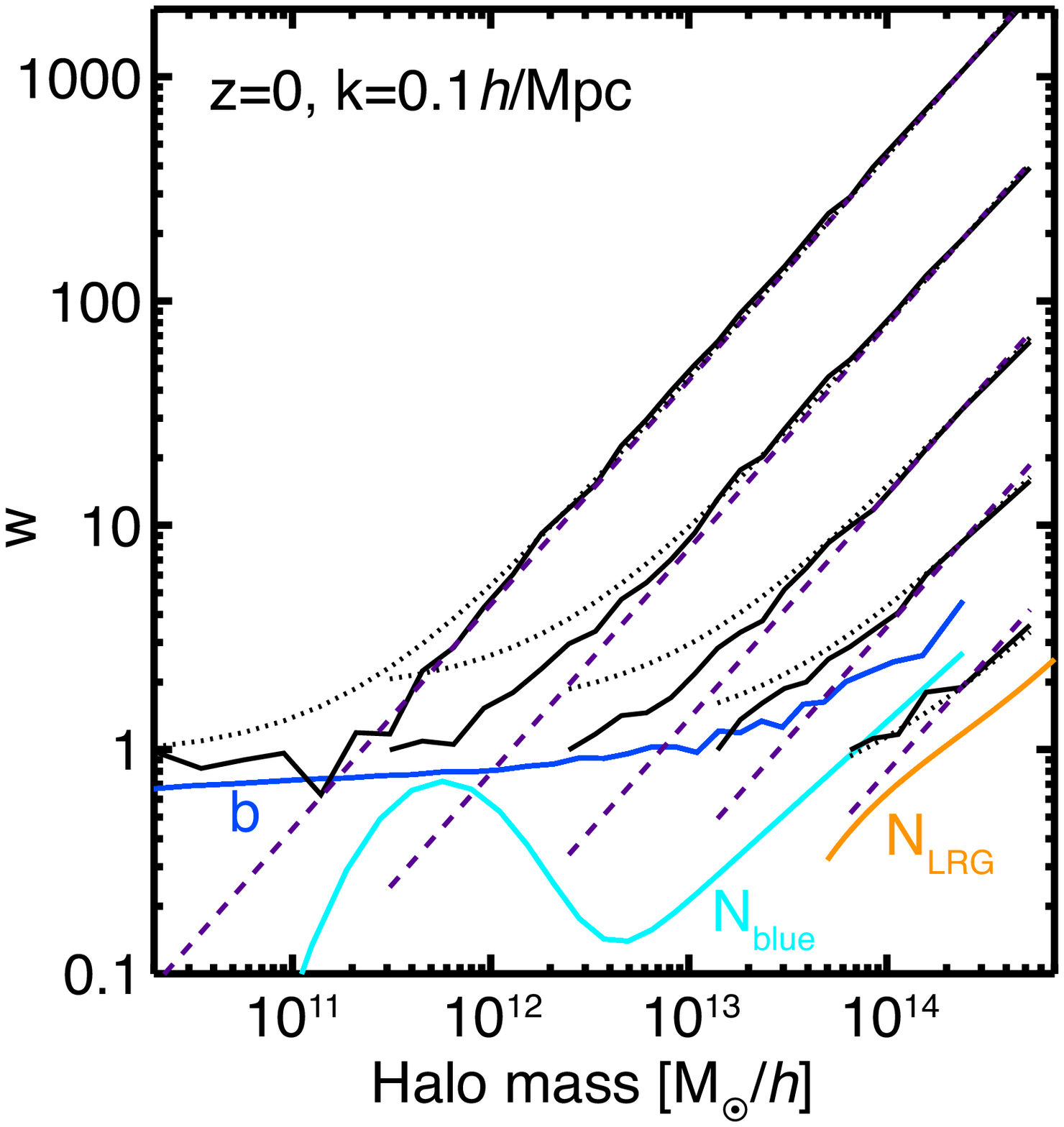}}
\end{center}
\caption{The weight function for optimal reconstruction of the mass 
  field on the scale $k=0.1h$Mpc$^{-1}$ at $z=0$ as a function of halo mass 
  for the NYU simulations (left) and the Millennium simulation (right).  
  The optimal weight depends on the minimum halo mass used in the 
 reconstruction; solid lines (with arbitrary vertical offset), which 
  are measured from simulations using Eq.~(14), show this dependence. 
  Dashed, dotted and blue curves show $w\propto M$, 
  equation~(\ref{halomodelw}), and $w\propto b$, 
  the latter being optimal if the Poisson model is correct.  
  The optimal weighting steepens as $M_{\rm min}$ decreases, approaching
  $w\propto M$, although not exactly along the dotted curves predicted
 by our halo model implementation of the sampling model.
  Cyan and orange lines show the mean halo occupancy
  distributions (HODs) for blue galaxies and luminous red
  galaxies, respectively, in the low-$z$ SDSS spectroscopic sample.
}
\label{wvsm}
\end{figure*}

\begin{figure*}
 \begin{center}
 \resizebox{\hsize}{!}{
  \includegraphics[angle=0]{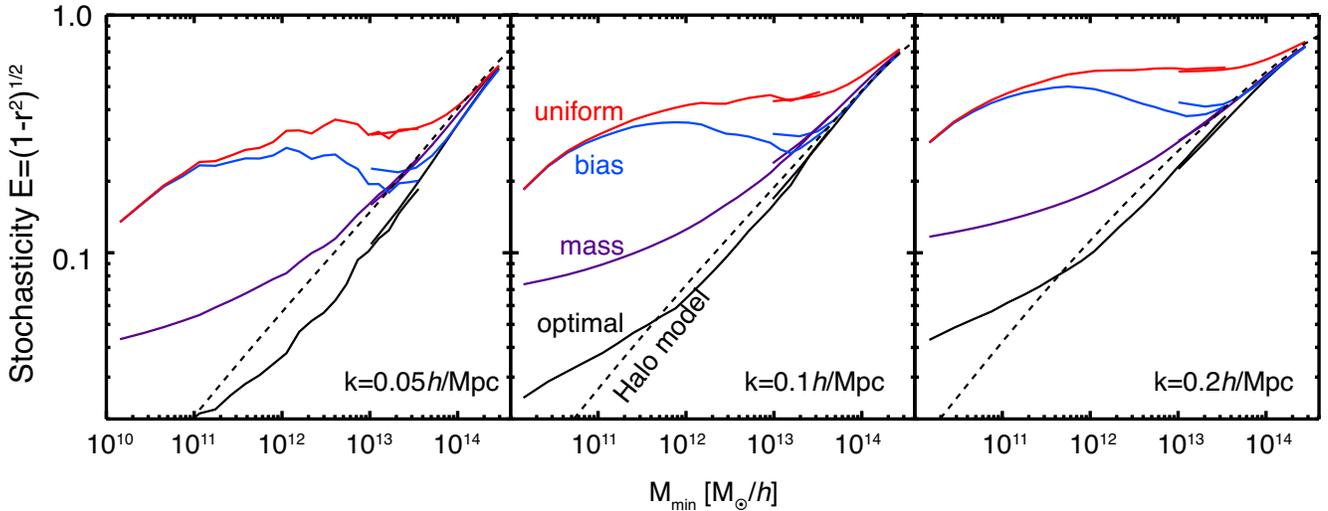}
 }
 \end{center}
 \caption{Stochasticity $E$ of the estimators of the mass field
  derived from the weights shown in Figure~\ref{wvsm}, shown
  as a function of the minimum mass of halos in the catalog. 
  The three panels show different $k$ values, all at $z=0$.
  In each panel, data at higher $M_{\rm min}$ are from the NYU
 simulations; lower $M_{\rm min}$ measurements are from
  the Millennium simulation.
  From the bottom up: black, purple, blue, and red solid curves show
  optimal, mass, bias and uniform weighting of the halos.
  Bias weighting would be optimal if the standard biased Poisson
  model were correct; it clearly is not.
  Dashed curve shows the halo model calculation of $E$ which
  assumes the mass is sum of halos that are
  Poisson-sampled from some halo field.  The failure of the
 model at $M<10^{12}h^{-1} M_\odot$ is discussed in the text.
}
\label{evsm}
\end{figure*}

\begin{figure*}
\begin{center}
\resizebox{\hsize}{!}{
\includegraphics[angle=0]{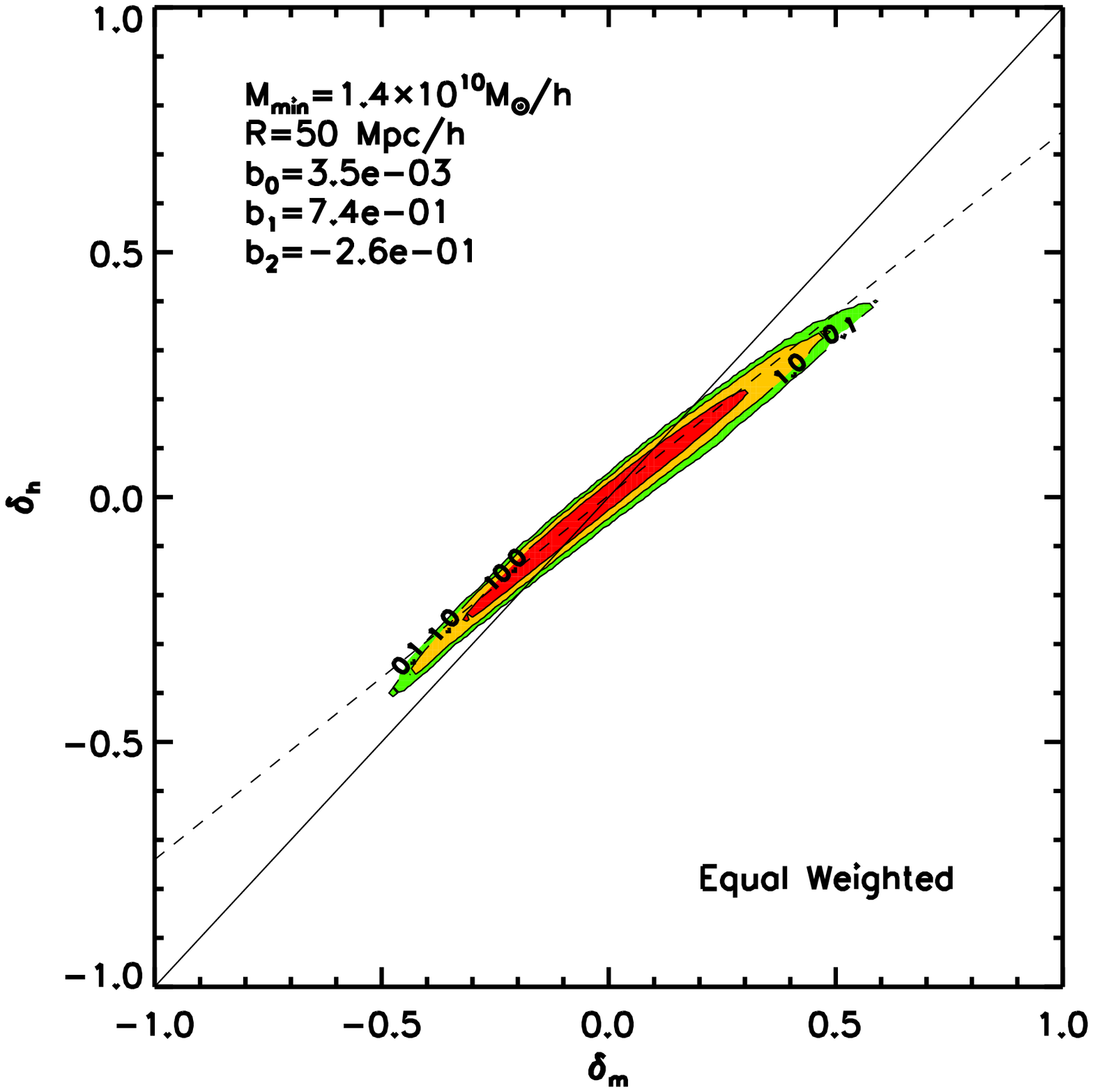}
\includegraphics[angle=0]{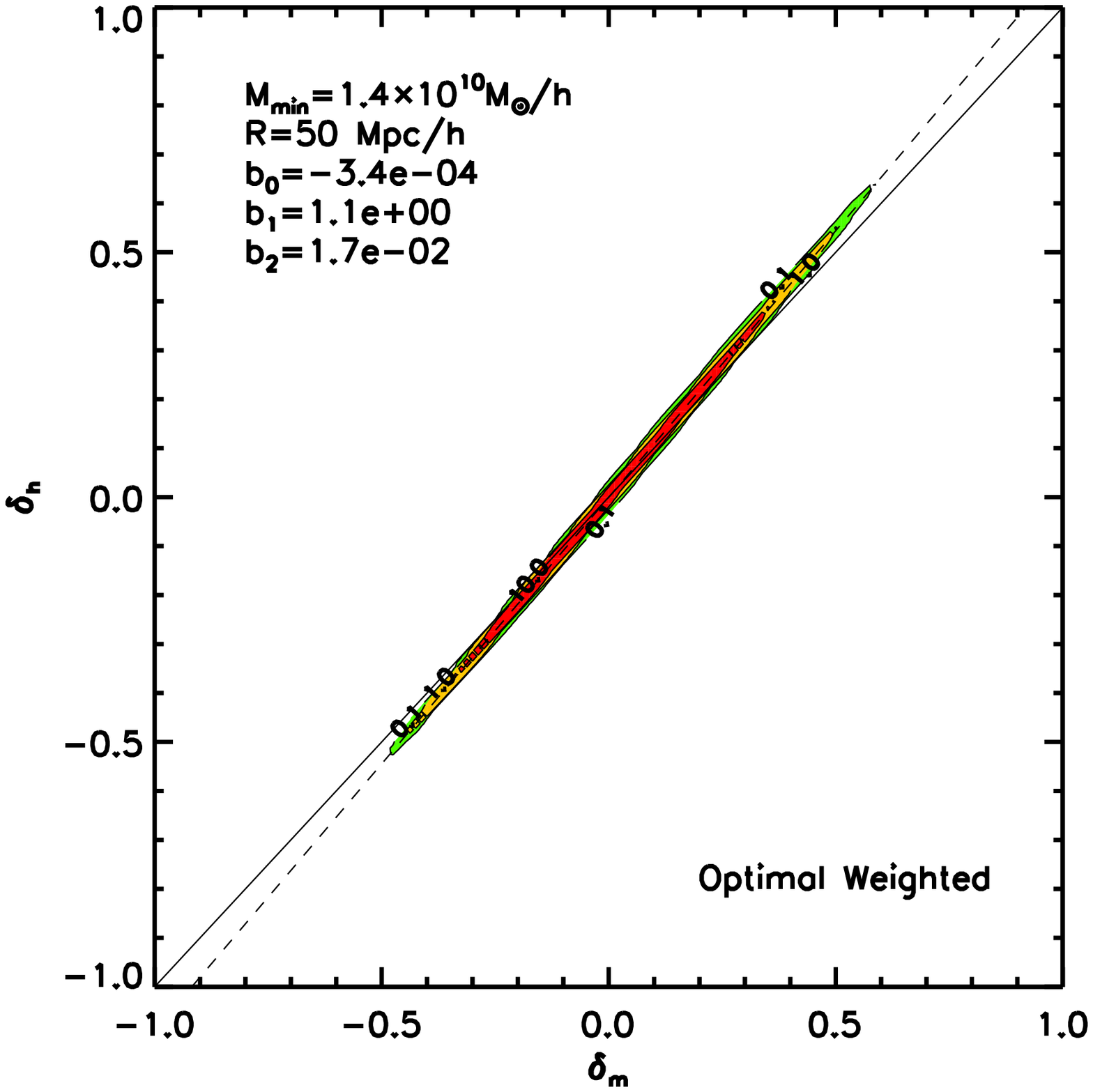}
}
\resizebox{\hsize}{!}{
\includegraphics[angle=0]{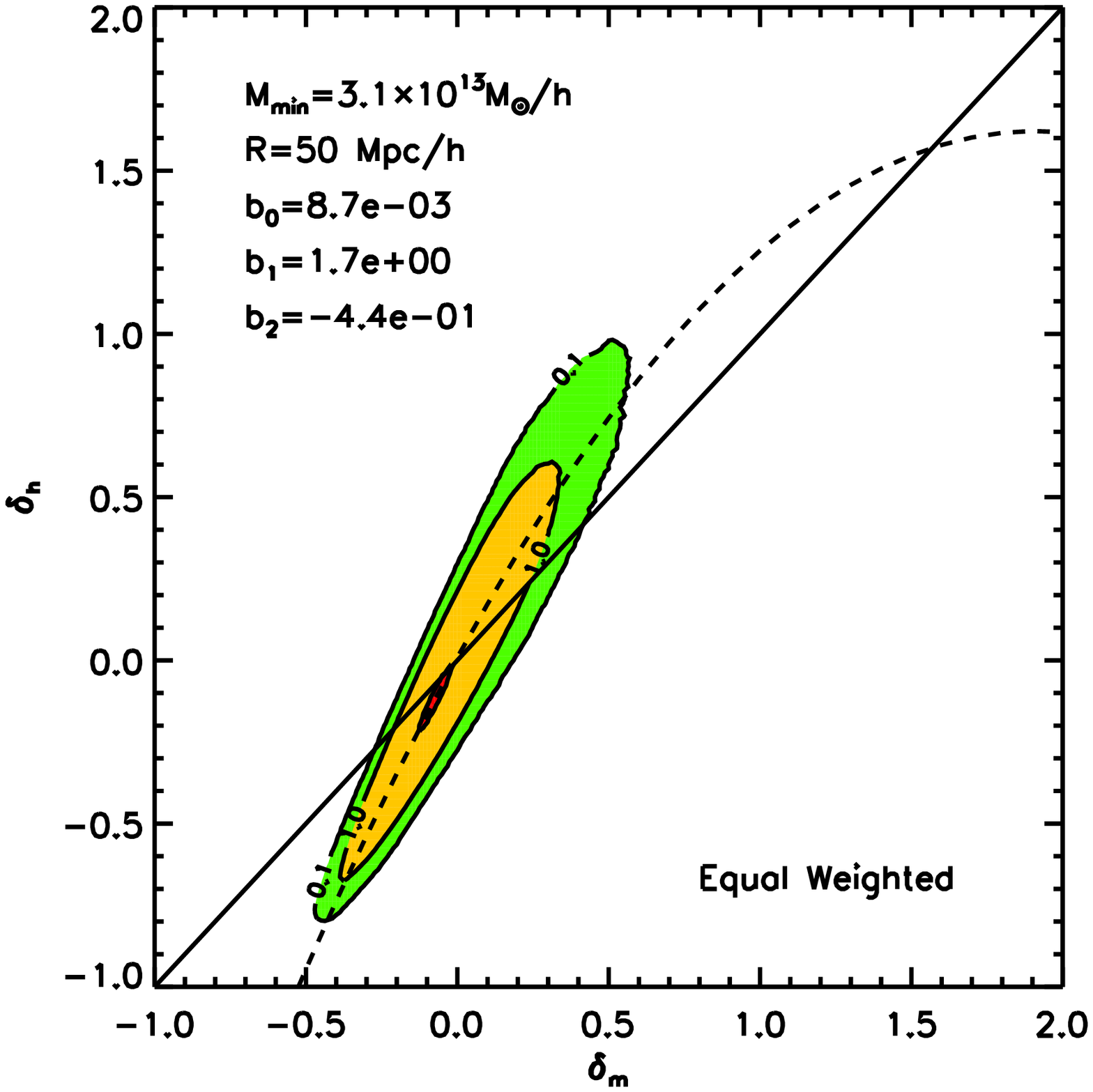}
\includegraphics[angle=0]{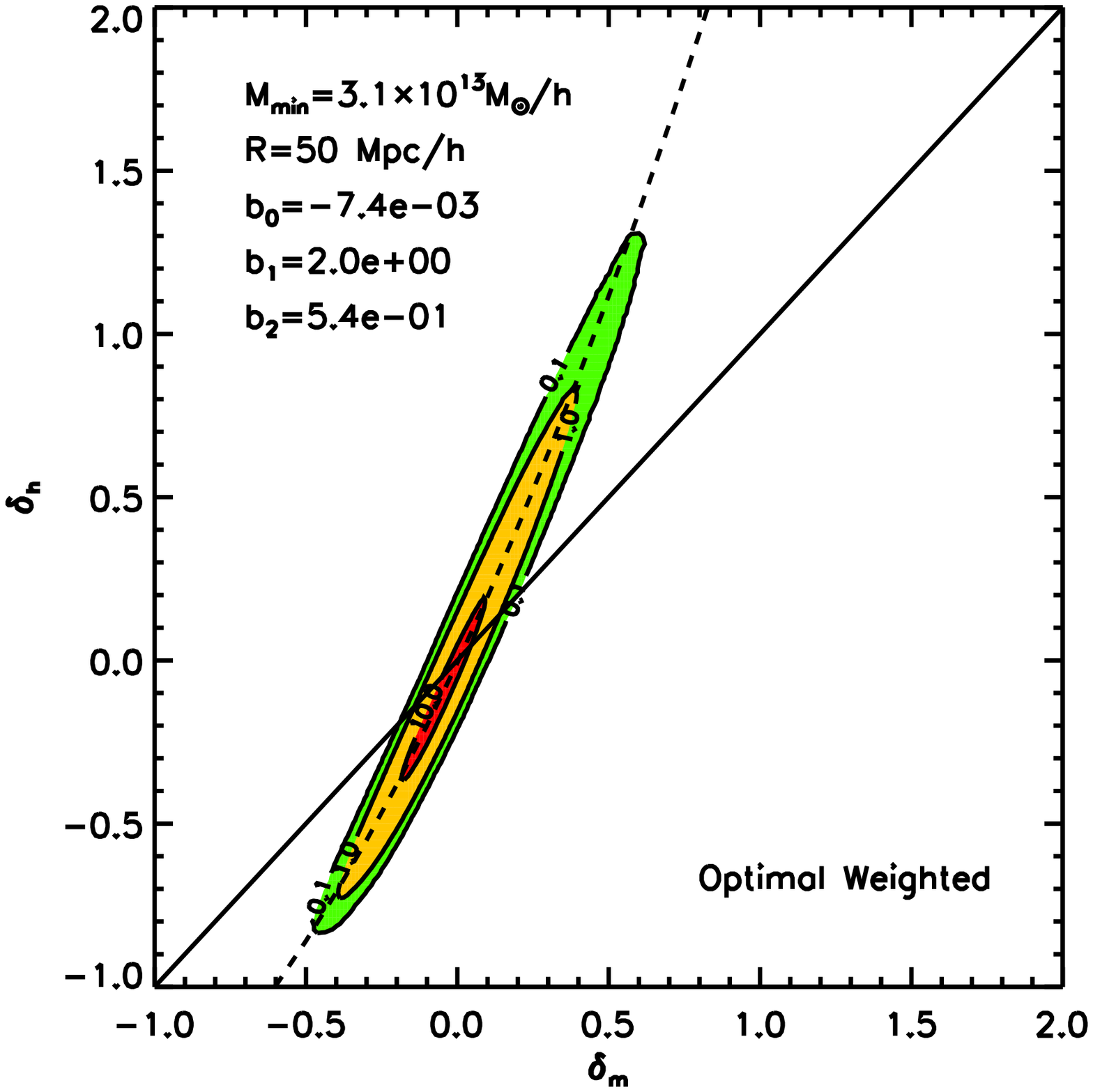}
}
\end{center}
\caption{Unweighted (left) and optimally weighted (right) real-space halo density 
fluctuations $\delta_h$ versus dark matter density fluctuations $\delta_m$ 
smoothed by a spherical top-hat window function with the radius of $R=50 h^{-1} Mph$. 
The three contour levels (0.1, 1, 10 shown in green, yellow, red) indicate the relative 
number density of data points in the $\delta_m$-$\delta_h$ plane.
Diagnal black solid lines indicate $\delta_h=\delta_m$. Dash lines show the 
fitting results of the function $\delta_h=b_0+b_1\delta_m+b_2\delta_m^2$, with 
best-fit parameters shown in the figures.
Top panels are results from the Millennium simulation with the 
mimimal halo mass of $M_{\rm min}=1.4\times10^{10}h^{-1}M_{\odot}$. Bottom 
panels show results from the NYU simulations, with $M_{\rm min}=3.1\times 10^{13}h^{-1}M_{\odot}$.} 
\label{Scatter}
\end{figure*}

\subsection{Measuring power spectra and covariance matrices in simulations}

We start our measurements by dividing halos into bins sorted by mass.
The clustering of halos depends weakly on the halo mass when halo mass
is low but increases rapidly with mass when $M>10^{13}~h^{-1}M_{\odot}$.
Furthermore halo abundances drop sharply at high mass.  As we do not
want a wide range of masses within a single bin, we include fewer
halos per bin at high masses.  

We choose bins so that the number of halos in each decreases exponentially
at high masses. The optimal weighting and stochasticity are robust to 
changes in the binning, as long as the function $b(M)$ is well sampled.  
We divide the halos into 10 bins for the NYU simulations, and use up 
to 30 bins for the Millennium simulation. We have tested that using 
more bins does not change our results.

Within each halo bin, we weight halos by their masses and assign them 
to a $N_g^3=256^3$ 3D mesh of cubic grid cells using the cloud-in-cell 
(CIC) assignment scheme \citep{Hockney81},  
i.e. we take the Fourier transform of the mass distribution within 
a halo bin.  This is in anticipation of the result below that optimal 
weighting is closer to mass-weighting than number-weighting of halos.  
If the bins are narrow in mass, this choice of intra-bin weighting 
should have little effect, which we have verified.

We separately Fourier transform the overdensity field of each halo bin 
and the total mass distribution. We correct each Fourier mode for the 
convolution with the CIC window function by the operation: 
\begin{equation}
\delta(\vk)=\delta(\vk)\,
  \left(\frac{\sin(x)}{x}\frac{\sin(y)}{y}\frac{\sin(z)}{z}\right)^{-2}, 
\end{equation}
where $\{x, y, z\}=\{k_xL_{box}/2N_g, k_yL_{box}/2N_g, k_zL_{box}/2N_g\}$, 
and $N_g$ is the number of grid cells in each dimension.
For each bin in $k$ we construct the covariance matrix of Fourier 
coefficients
 $C_{ij}(k)=\langle\delta_i(k)\,\delta_j(k)\rangle$,  
where $i$ and $j$ range over all halo bins, 
as well as the mass power $P=\langle \delta_m^2\rangle$, and hence 
the covariance biases \vb\ of the halos against the mass.
For the NYU simulations we average results from all 49 realizations to
produce a mean covariance matrix.

\begin{figure*}
\begin{center}
\resizebox{\hsize}{!}{
\includegraphics[angle=0]{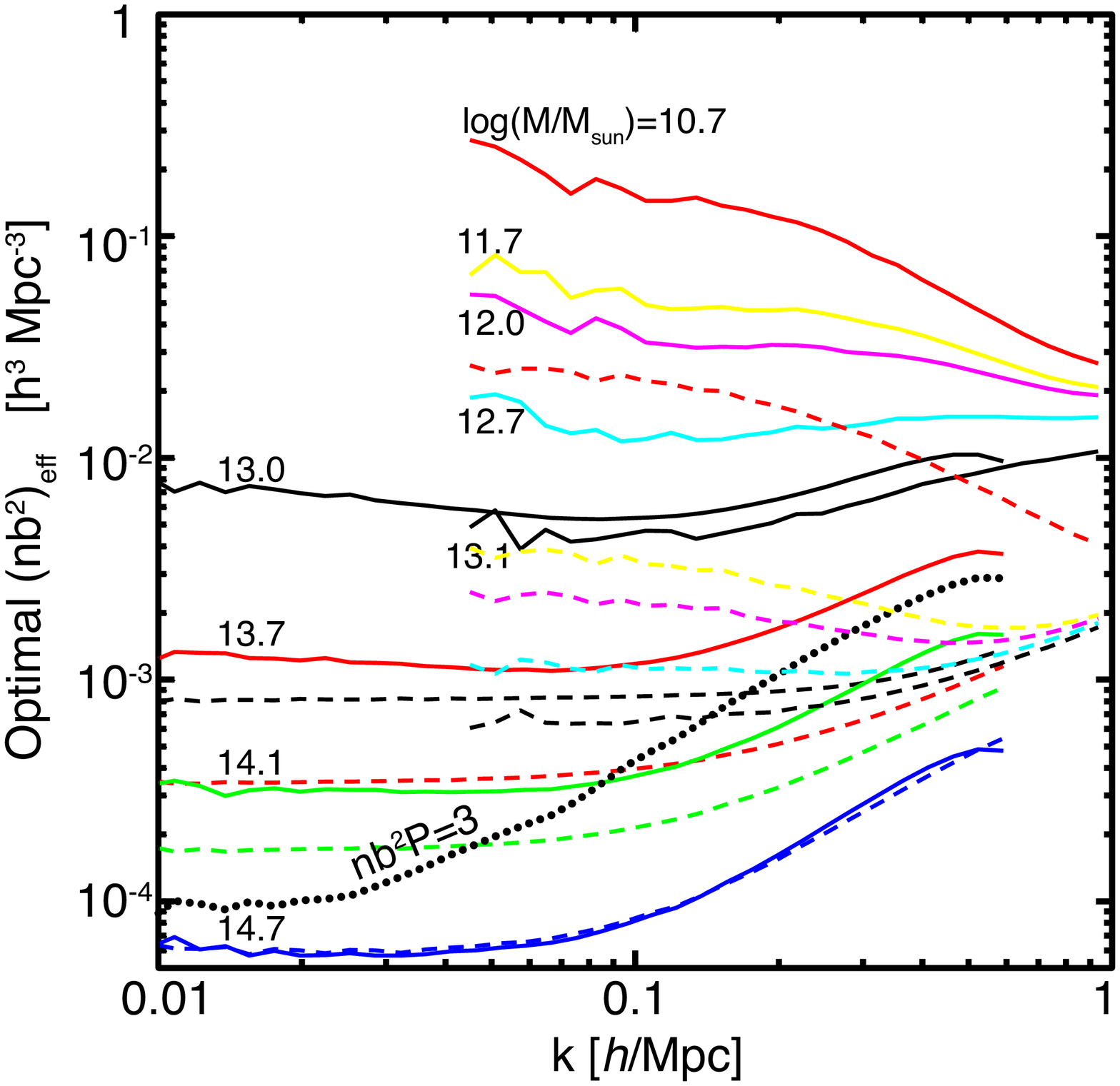}
\includegraphics[angle=0]{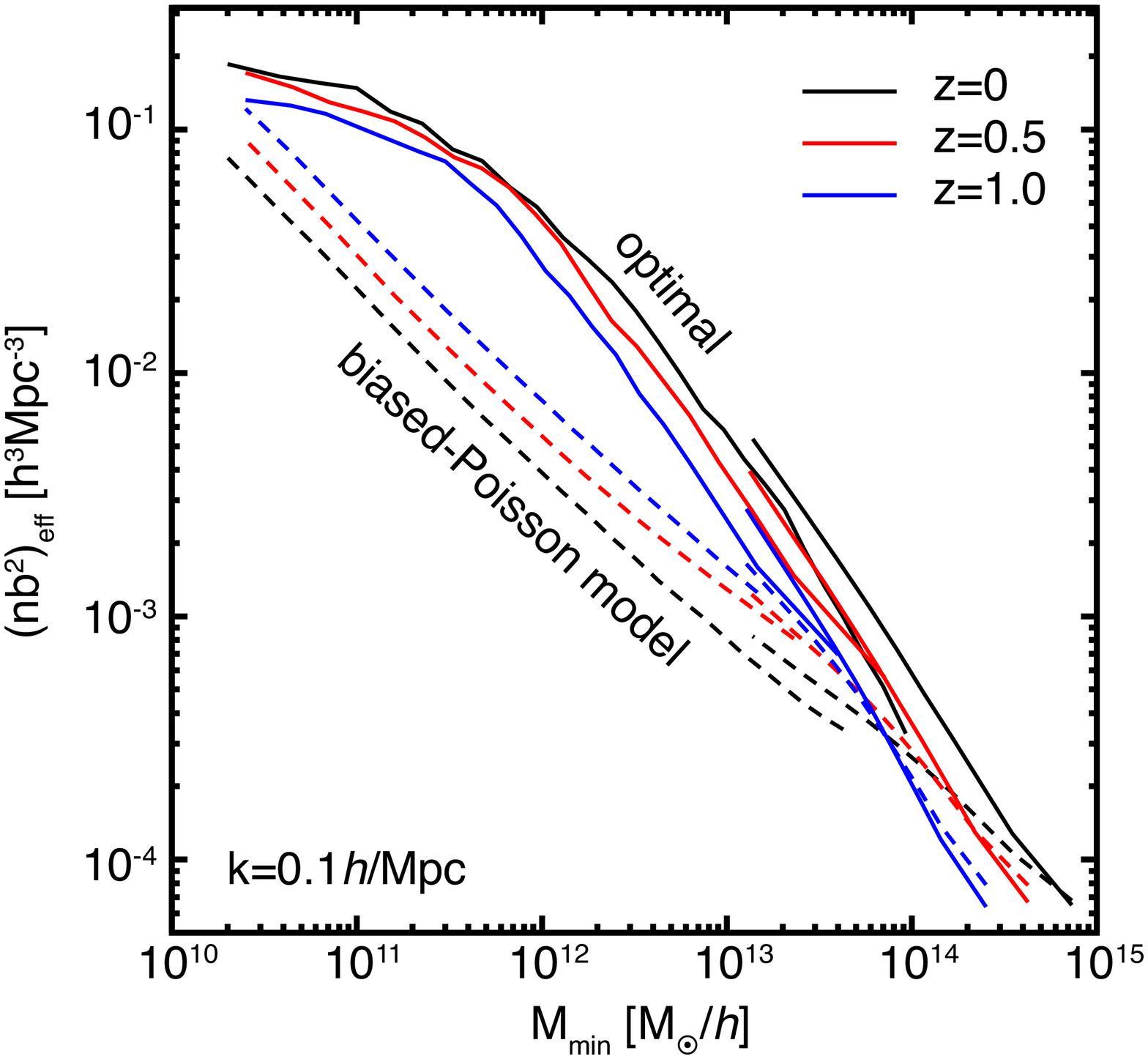}
}
\end{center}
\caption{{\em Left:} The effective source density $(nb^2)_{\rm eff}$
  of an optimally-weighted mass estimator is plotted (solid) vs $k$ for
  catalogs with a different halo mass cutoffs $M_{\rm min}$ as
 labeled.  This measure of stochasticity is found to vary little
  with scale in the linear regime $k<0.1h$Mpc$^{-1}$. The dashed lines show
  the $nb^2$ expected via Equation~(\ref{nbpoi}) under a biased-Poisson
  model of galaxy stochasticity.  The dotted line shows
  the $(nb^2)_{\rm eff}$ required to attain $E=0.5$, {\it i.e.} a
  volume-limited power spectrum measurement.
 {\em Right:} 
  For three different redshifts, we plot the optimal $(nb^2)_{\rm
    eff}$ vs the minimum mass of the halo catalog used for mass
  reconstruction (solid lines).  The dashed lines plot the $nb^2$ that
  we would expect if halos were a Poisson sampling of the mass
  distribution, as per Equation~(\ref{nbpoi}).  Because halos comprise
  the mass rather than sampling the mass, the $(nb^2)_{\rm eff}$ is up
  to $15\times$ higher, {\it i.e.} the mass estimator is less
  stochastic, than the Poisson model predicts.
}
\label{nb2vsk}
\end{figure*}

\subsection{Measured stochasticity and optimal weights}
\label{measure}

Figures~\ref{wvsm} shows the optimal weights we derive from the 
simulations (black solid lines), and compares them with various 
functions of mass.  Purple-dashed lines show $w\propto m$;
blue curves show the bias weighting $w\propto b$ that would
be appropriate if the standard biased-Poisson model were correct, 
and dotted curves show the optimal weight in equation~(\ref{halomodelw})
derived from the halo model of Section~\ref{halomodel}.  

Clearly, neither $M$ nor $b$ are optimal weights.  
Indeed, the shape of $w_{\rm opt}(M)$ depends on the cut-off mass 
$M_{\rm min}$ of the halo catalogue.  (This dependence follows that
found by H10, in their study of $\sigma_w$.) 
When $M_{\rm min}<10^{13}~h^{-1}M_{\odot}$, the massive end of the
$w_{\rm opt}(M)$ is close to mass weighting, as illustrated by the 
right-hand plot of Figures~\ref{wvsm}. 
When $M$ is close to $M_{\rm min}$, however, $w_{\rm opt}(M)$ is
flatter than mass weighting. Moreover, the slope of $w_{\rm opt}(M)$ 
gets shallower as $M_{\rm min}$ increases, as shown in the left hand 
plot. Weighting halos by their masses is a poorer approximation to the
optimal weight when $M_{\rm min} >10^{13}~h^{-1}M_{\odot}$.
The halo model prediction of the optimal weight is generally in good 
agreement with the measurements.  The agreement is not perfect, however, 
especially when $M$ approaches $M_{\rm min}$. 

Figure~\ref{evsm} shows the stochasticity $E$ associated with these
linear estimators $\hat \delta_m$ of the mass distribution, as a
function of the minimum mass $M_{\rm min}$ of halos included in the
sample. Black, purple, blue, and red solid curves show $E$ derived from
optimal, mass, bias and uniform weighting of the halos. 
Weighting halos by their masses yields lower $E$ than bias weighting 
or equal weighting, but is significantly worse than the optimal when 
the halo catalog has $M_{\rm min}\approx 10^{13}h^{-1}M_\odot$ or lower.
Bias weighting would be optimal if the standard biased Poisson model 
were correct, but is clearly far from optimal for halos in $N$-body 
simulations.  
The dashed curves show the halo model description of $E_{\rm opt}$ 
(equation~\ref{halomodelE}).  The model agrees with the measurements 
at $M_{\rm min}>10^{12}h^{-1}M_{\odot}$, but it does not predict the 
inflection we measure at smaller $M_{\rm min}$. 
We discuss this further in Section~\ref{depart}.

As an additional check of our numerical methods, we have verified 
that inclusion of the ``dust bin''  in the estimator leads to a 
perfect mass estimator ($E=0$) with optimal weights directly 
proportional to halo mass.  

\subsection{The scatter between the halo field and the mass field}
To illustrate the gain from applying the optimal weights, we show in 
Figure~\ref{Scatter} the scatter between the fluctuations of the 
halo field $\delta_m$ and the mass density field $\delta_m$, 
before and after applying the optimal weights. Notice that 
$\delta_h$ and $\delta_m$ are both density contrasts in 
configuration space that are smoothed by the same spherical 
top-hat window function. We do the smoothing by multiplying
the density contrasts in Fourier space with the Fourier transform 
of the window function, $\delta_{h,m}(k)=W_R(k)\delta_{h,m}(k)$, where 
$W_R(k)=3[\sin (kR)-kR\cos (kR)]/(kR)^3$ and $R$ is the radius of the window function.
Then we Fourier transform back and get the smoothed $\delta_h$ and $\delta_m$.

We fit the scatter plots with the 
polynomial function $\delta_h=b_0+b_1\delta_m+b_2\delta_m^2$, 
to see if there is any indication of non-linear bias factor $b_2$. 
We usually find very small fitted values of $b_2$, especially for the 
optimal weighted cases. We also find an increase of $b_2$ 
value when increasing the low mass cut of the halo sample. 
In general, we see a significant improvment of 
applying the optimal weights, indicated by the shrinking of the scatter. 
This shows that the optimal weights indeed work well, without any 
higher-order bias correction.

\subsection{Scale dependence of stochasticity}
Figure~\ref{nb2vsk} illustrates that the optimal $(nb^2)_{\rm eff}$ is
nearly independent of $k$ at fixed $M_{\rm min}$ in the linear
regime, where $(nb^2)_{\rm eff}$ is related to $E_{\rm opt}$ by 
equation~(\ref{nb2eff}).  Since both $(nb^2)_{\rm eff}$ and the 
Poisson prediction $(nb^2)$ (the simple bias weighted sum over 
the halo population), are nearly constant across the linear regime, 
we compare them in the right panel of Figure~\ref{nb2vsk} 
as a scale-independent measure of stochasticity.
We find that at all redshifts and $M_{\rm min}$ values, the achievable
$(nb^2)_{\rm eff}$ is significantly better (higher) than would have
been expected in the model where halos are a Poisson sampling of the
mass.  The ratio $(nb^2)_{\rm eff}/(nb^2)$ can be as high as $\approx 15$ 
for surveys of $M > 10^{12}h^{-1}M_\odot$ halos at $z=0$.  
Even for surveys limited to massive clusters, 
$M_{\rm min}\approx 10^{14}h^{-1}M_\odot$, the effective source
density of the optimal estimator is $\approx 2\times$ better than the
Poisson model predicts. 

\subsection{Departures from the halo model}
\label{depart}

The dashed curve in Figure~\ref{evsm} plots the halo model prediction
of the optimal $E$, which assumes the mass is comprised of halos that 
are Poisson-sampled from some halo field.
The $E_{\rm opt}$ from simulations becomes shallower for $M_{\rm
  min}<10^{12}h^{-1}M_\odot$ where as the halo model does not.
Either the halo model is not accurate at 
$M_{\rm min} < 10^{12}h^{-1}M_\odot$, or there is some bias in the
numerical estimation of $E_{\rm opt}$ from the simulation catalogs. 

Calculating $E_{\rm opt}^2$ at $M_{\rm min} < 10^{12} h^{-1}M_\odot$ sets 
heavy demands on the measurement of the covariances in the simulation,
because $\vb^T \mC^{-1} \vb$ must be calculated to a fractional
accuracy of $E^2_{\rm opt} < 10^{-3}$ in this regime.  
We have considered the possibility that $E_{\rm opt}$ levels off at 
small $M_{\rm min}$ because of discreteness effects.  Specifically, 
in the simulations, mass comes in units of $m_p$, so the ``dust'' is 
not made of arbitrarily small halos.  This makes 
$\noise_d\to \noise_d + 1/n_d$ approximately.  However, this 
additional factor is too small to explain the flattening we see.  We
have also verified that the plateau in $E_{\rm opt}$ is unaffected by
the size of the bins in mass or $k$.

\begin{figure}
\begin{center}
\resizebox{\hsize}{!}{
\includegraphics[angle=0]{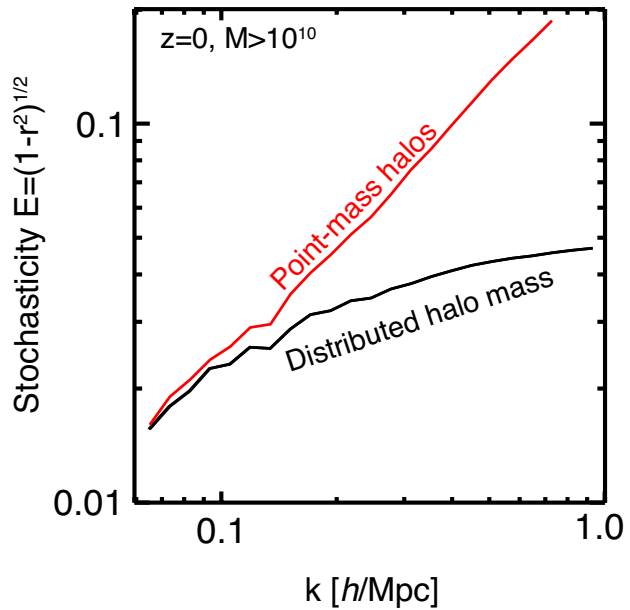}
}
\end{center}
\caption{The optimal stochasticity $E$ as a function of 
wave-number $k$ from using all halos in the Millennium simulation is
shown for two cases: the upper red curve treats each halo as a point
mass, while the lower black curve uses the full spatial distribution
of the particles comprising the mass of each halo.  If the halo
catalog does not contain information on the variability of internal
halo structure at a given mass, it cannot fully map the mass
distribution.  This significantly degrades $E$ for $k>0.15h\,{\rm
  Mpc}^{-1}$, but does not explain the inflection in $E_{\rm opt}$
observed at $k\le 0.1h\,{\rm Mpc}^{-1}$ for $M<10^{12}h^{-1}M_\odot$
in Figure~\ref{evsm}.
}
\label{fullHalos}
\end{figure}

As noted earlier, the stochasticity $E$ will be degraded if halos of a
given mass have varying internal structure, while we treat halos 
of a given mass as being identical point masses in the analysis.  
Within the halo model, the stochasticity adds to $\noise_m$ as per 
(\ref{structurenoise}).
We test the magnitude of this effect by creating new halo overdensity 
maps from the full sample of $N$-body particles belonging to halos in 
each mass bin.  These ``true'' halo mass maps are then used to create 
an optimal mass estimator.  Figure~\ref{fullHalos}  shows
that the point-mass approximation has negligible impact on $E$ at
$k<0.15h\,{\rm Mpc}^{-1}$, but at higher $k$ the mass estimator is
increasingly degraded by the absence of information on 
the variability of internal structure of massive halos.

Even when we use the full halo mass distributions in our optimal
estimator, the performance for $M_{\rm min}<10^{12} M_\odot$ remains
worse than $E$ predicted by the halo model.
It is possible that noise in \vb\ and \mC\  from having 
too few modes is compromising our measurements of $E_{\rm opt}$ in the
Millennium simulation.  This would most strongly affect the
low-$k$ region where modes are scarcest, however the inflection
does not exhibit this behavior.  

We conclude 
that we have reached limits of the
assumption that halos are a Poisson sampling of an underlying halo 
field.   For $M \approx 10^{12}h^{-1}M_\odot$,  $\eta_d \approx 0.66$ 
so the effects of exclusion/mass conservation are beginning to 
matter, so it is perhaps not surprising that the optimal mass
reconstruction is poorly described by a model that presumes halos to
be independently sampled from the halo field.  We will defer to future
work investigation of mass reconstruction in the presence of exclusion
and other non-linear effects.

\subsection{Explicit test of the sampling model}
\label{testsampling}
\begin{figure}
\resizebox{\hsize}{!}{
\includegraphics[angle=0]{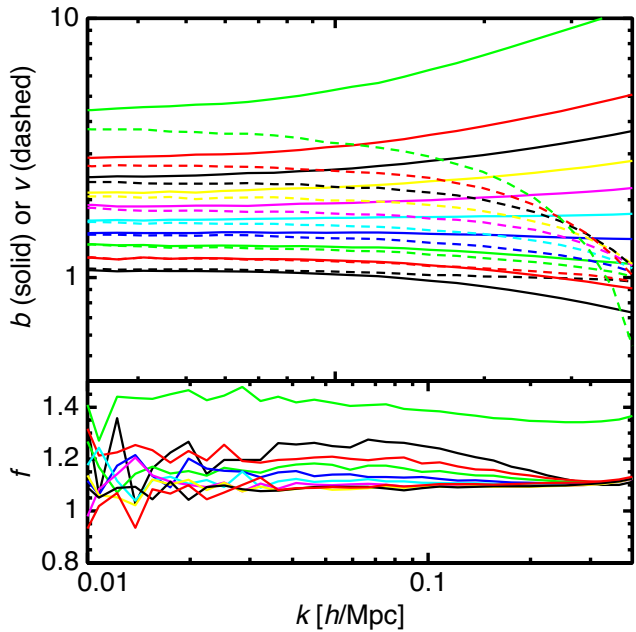}
}
\caption{The best fit of the sampling model
  $C_{ij}=v_iv_jP + \delta_{ij} f_i/n_i$ to the $z=0$ halo catalog 
  of the NYU simulation is shown as a function of $k$.  The solid 
  lines in the upper panel plot the bias $b_i$ of halos in each mass 
  bin, while the dashed lines are the best-fit $v_i$ for each bin.  
  Higher-mass bins have higher bias.  Note that $v_i < b_i$ for
  massive halos, by an amount that grows at $k>0.1h\,{\rm Mpc}^{-1}$, 
  but that $v_i > b_i$ at lower masses, as predicted.
  The lower panel shows the best-fit $f_i$.  Poisson sampling would 
  induce values slightly above $f_i=1$ because of the mass weighting 
  within a halo bin, particularly for the most massive bin, as observed.
}
\label{bvn}
\end{figure}

Are the observed covariance matrices of the halos consistent with
their being stochastic discrete realizations of an underlying ``halo
field'' $\delta_h$, the model of \S\ref{samplingmodel}?  We answer
this question by asking how well the elements $C_{ij}$ of the
simulations' covariance matrices can be fit by appropriate values of
the $v_i$ and $f_i$.  We find the $\{v_i,f_i\}$ which minimize the L2
norm of the residual to the model (\ref{vvNmodel}):
\begin{equation}
\lVert\delta \mC\rVert_2 \equiv \sum_{ij} \left( v_i v_j +
  \delta_{ij}f_i/n_iP - C_{ij}/P\right)^2.
\end{equation}
(We use the mass power $P$ instead of $P_h$ in the fit, which slightly
changes the values of $v_i$ and $f_i$ without affecting the quality of
the fit.)
Figure~\ref{bvn} plots the best-fitting values of $f_i$ and $v_i$ for 
each halo mass bin in the NYU simulations, along with the bias $b_i$.  
The fitted values of $f_i$ slightly exceed unity, but this is consistent
with halos being a Poisson sampling ($f_i=1$) of a ``halo field''
because the mass-weighting within each halo bin will cause $f_i$ to
rise slightly above unity, particularly in the most massive bin.

The fit to the sampling model does confirm a departure from the
simple biased-Poisson model, however, in that the biases $v_i$ of the 
halos with respect to the parent halo field are not equal to the 
covariance biases $b_i$ with respect to the mass field.  There is a 
divergence of \vb\ from \vv\ toward the trans-linear regime.

\begin{figure}
\resizebox{\hsize}{!}{
\includegraphics[angle=0]{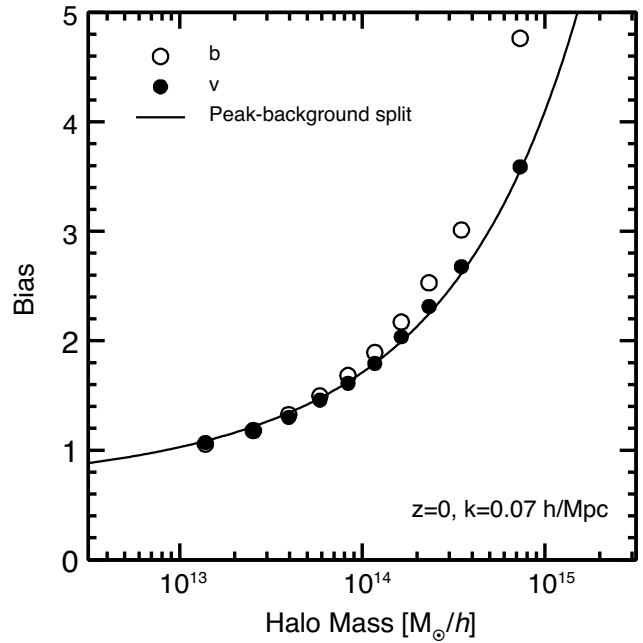}
}
\caption{For the full NYU halo sample at $z=0$, we plot three
  types of bias.  The open symbols are the covariance bias $b$ vs the
  mass distribution.  The solid symbols are the bias $v$ vs the
  underlying continuum ``halo field'' $\delta_h$ in the sampling model
  that best fits the halo-mass covariance matrix.  The curve is the
  analytic bias prediction of the peak-background split model.
}
\label{pbs}
\end{figure}

The peak-background split model yields an analytic prediction 
$b_{\rm pbs}$ for the bias of halos vs the underlying mass distribution.
Figure~\ref{pbs} shows that for the NYU simulations, it is the bias
$v$ of the halos vs the {\em halo field} $\delta_h$ that is best
described by $b_{\rm pbs}$, not the bias $b$ of the halos vs the 
{\em mass distribution.}  
This observation should lead to a deeper theoretical understanding 
of the halo distribution.  In particular, it may help resolve the 
discrepancy reported by \cite{Manera10} between $b_{\rm pbs}$ and 
their measurements of halo bias, which were effectively what we 
call $b$, rather than $v$.

The quality of the fit of the sampling model to the covariance matrix
can be gauged by the ratio
\begin{equation}
R \equiv \frac{\sqrt{\lVert\delta \mC\rVert }}{\lVert \vv \rVert}.
\end{equation}
This quantity measures the ratio of the RMS error in elements of \mC\  to
the RMS value of the model---excluding the diagonal elements, which
are always perfectly fit by adjusting the $f_i$.  
For the NYU
simulations at $z=0$ we find \mC\ well fit by this model:
$0.01<R<0.035$ for all $k<0.25~h$Mpc$^{-1}$.
For the bin at
$k=0.07~h$Mpc$^{-1}$ in the NYU simulations, we would expect
$R\approx0.02$ from Gaussian sample variance in the estimation of the
$C_{ij}$.  At higher $k$, the sample variance in the estimation of
\mC\ should drop, so it is likely that by $k=0.2$ the residuals of
\mC\ to the sampling model are in excess of statistical errors.  We
expect effects such as halo exclusion and halo internal structure
to induce departures from the simplest sampling models at scales
approaching the halo sizes.

The sampling model is not able to fit the $\mC$ matrix across the full
mass range of the $z=0$ Millennium halo catalog, yielding $R>0.1$.
Excluding the two most massive of 30 halo bins from $\mC$ permits a
solution with $R\approx0.07$ in the linear regime, however this
solution requires values of $f_i<1$ or even negative $f_i$ for the
least massive halos.  We take this as an additional sign that low-mass
halos cannot be considered to populate the halo field independently,
e.g. exclusion may be important.  The much smaller volume of the
Millennium simulation leads to larger statistical
fluctuations in the elements of \mC, so at this time we refrain from
detailed analysis of departures from the sampling model for
less-massive halos.

\section{Practical consequences}
\label{andso}
With an optimal (linear) mass reconstruction algorithm in hand, we can
examine strategies for conducting cosmological measurements with
maximal efficiency.

\subsection{Power spectrum measurement with halo mass estimates}
We first posit a survey attempting to measure the shape of the matter
power spectrum near $k=0.2~h$Mpc$^{-1}$, {\it e.g.} a baryon acoustic
oscillation measurement. Since the error in a power spectrum
determination from $N_m$ modes of a stochastic estimator is
$\sigma_P/P = [(1-E^2)N_m/2]^{-1/2}$,  it is typical to consider a
survey with $E\le 0.5$ as sample-variance limited.  This is equivalent
to the criterion $(nb^2)_{\rm eff}P = 3$.  [Here we assume that the
bias of the estimator is known or immaterial.]

If the redshift of a halo can be obtained with a single spectrum of 
its central galaxy, then clearly the strategy for attaining a mass
estimator with a given stochasticity with the fewest redshift 
measurements is to measure redshifts for all halos above a chosen mass limit.
In practice one could identify candidate halos from multicolor imaging
data using a variant of red-sequence detection or cluster-finding with
photometric redshifts \citep[e.g.][]{MaxBCG,RCS}.

If imaging data can be used successfully to identify clusters and 
estimate their host halo masses, then the more expensive spectroscopic 
redshift survey need target only the brightest member of each cluster 
or group to determine a redshift for the presumed dark matter halo. 
X-ray or Sunyaev-Zeldovich survey data could also potentially contribute to 
halo finding and mass estimation.  

What is the minimum number of spectra that one must obtain to make the
volume-limited power spectrum measurement described above? 
We answer this question for the case when halos are 
optimally weighted, and also for comparison the (incorrect) prediction
for halos that occupy the mass by the biased-Poisson model.
We list in Table~\ref{tab2} 
the minimal mass of halos and minimal number density of halos for the 
above two cases in three different redshifts. The optimally weighted
case needs a factor 4--12 fewer spectra than predicted by the
biased-Poisson model to achieve the sample-variance limit $E\le0.5$.

To complete a volume-limited survey for $0<z<1$ 
that achieve $(nb^2)_{\rm eff}P = 3$ at the BAO scale 
$k\sim 0.2~h$Mpc$^{-1}$, the total number 
of spectra one needs is $\approx6\times 10^6 f_{\rm sky}$ if halos are optimal weighted. 
If the biased-Poisson model were correct, one would require
$\approx 33\times10^6 f_{\rm sky}$, a factor of 5 more. 
For a redshift survey to $z<0.7$ such as the ongoing SDSS-III
Baryon Oscillation Spectroscopic Survey (BOSS)\footnote{http://www.sdss3.org/cosmology.php},
the two cases yield $2\times10^6 f_{\rm sky}$ and $14\times10^6 f_{\rm
  sky}$, respectively.  Hence BOSS at $f_{\rm sky}=0.25$ would require
only 500,000 optimally targeted and weighted redshifts to achieve
$(nb^2)_{\rm eff}P = 3$, while the survey plans to obtain 1.5~million
redshifts.   Given that precise halo masses are not
easily accessible, the weighting for a real survey may be sub-optimal.
We will show in Section~\ref{LRG} that if one weights halos by the 
numbers of LRGs, a factor of 3 more redshifts are required to attain $E<0.5$.

\begin{table*}
\centering
\caption{The number density of halos $n^{\rm opt}$ and the corresponding minimal halo
mass $M^{\rm opt}_{\rm min}$ needed to achieve $(nb)_{\rm eff}^2P=3$
($E=0.5$) at BAO scales ($k=0.2h\,{\rm Mpc}^{-1}$) when applying the optimal weighting.
For comparison, the minimal halo mass $M^{\rm P}_{\rm min}$ and number density $n^{\rm P}$ needed to achieve 
the same accuracy in a Poisson model are also listed.  The last column
gives the ratio of required redshift measurements in the two models.}
\bigskip
\begin{tabular}{lccccccc}
\hline
\hline
$z$  &$M^{\rm opt}_{\rm min}$ & $M^{\rm P}_{\rm min}$ & $n^{\rm opt}$ & $n^{\rm P}$  & $n^{P}/n^{opt}$  \\
&$h^{-1}M_{\odot}$&$h^{-1}M_{\odot}$&($h~$Mpc)$^{-3}$& ($h~$Mpc)$^{-3}$ &  \\
\hline
0.0 & $6.2\times 10^{13}$ &$6.5\times 10^{12}$& $4.2\times10^{-5}$ & $5.7\times10^{-4}$ & 13 \\
0.5 & $2.6\times 10^{13}$ &$5.2\times 10^{12}$& $8.0\times10^{-5}$ & $5.7\times10^{-4}$ & 7  \\ 
1.0 & $1.1\times 10^{13}$ &$3.8\times 10^{12}$& $1.4\times10^{-4}$ & $5.8\times10^{-4}$ & 4 \\ 
\hline
\hline 
\end{tabular}
\label{tab2}
\end{table*}

\subsection{Bias calibration with weak lensing}
A more ambitious measurement is to cross-correlate the mass distribution
estimated from a galaxy redshift survey with a weak gravitational
lensing shear map, thereby calibrating the bias of the estimator
\citep{Pen04}.  Measures of the redshift dependence of the
cross-correlation between lensing and matter can also strongly
constrain the curvature and $D(z)$ function of the Universe
\citep{Bernstein04}.  A simplified analysis of these problems considers
the covariance matrix between the gravitational convergence field
$\kappa$ and the galaxy-based mass estimator $g$ to be
\begin{equation}
\label{ckg}
\mC = \left(\begin{array}{cc}
P + \noise_\kappa & bP \\
bP & b^2P + \noise_g
\end{array} \right).
\end{equation}
where $\noise_\kappa$ is the noise in the weak lensing mass estimation.
To fully exploit the lensing data, the mass estimator should attain
$E^{-2}=1+b^2P/\noise_g \ge P/\noise_\kappa$ such that the $S/N$ ratio per mode
  of the mass estimator exceeds the $S/N$ ratio per mode of the
  lensing map.

If the lensing noise level $\noise_\kappa$ is known and one is 
inferring $P, b,$ and $\noise_g$ from the values of the lensing and mass
estimators in $N_m$ modes of the sky, then the marginalized error in
the estimate of $b$ becomes
\begin{equation}
\label{sigbxc}
\frac{\sigma_b}{b} = \sqrt\frac{ (1+\noise_\kappa/P) \left( 1/(nb^2)_{\rm eff}P + \noise_\kappa/P\right) +
  (\noise_\kappa/P)^2}{N_m}.
\end{equation}
When the lensing and mass reconstruction both have high $S/N$ per
mode, this becomes
\begin{equation}
\frac{\sigma_b}{b} 
\approx \sqrt\frac{\noise_\kappa/P + E^2}{N_m}.
\end{equation}
As an example, consider a lensing source plane consisting of $n=30$
galaxies arcmin$^{-2}$ as $z_s=1$ being cross-correlated with a
transverse mass mode at $k=0.1h/$Mpc at $z=0.5$, near the peak of the
lensing kernel.  The shear signal will appear at multipole
$\ell=kD\approx130$ where the total power in the shear signal is
$C_\ell \approx 8\times10^{-8}$.  The shape noise power is
$\sigma^2_\gamma / n \approx 2\times10^{-10}$, giving 
$P/\noise_\kappa\approx400$, or a $S/N$ per mode
of $\approx 20$.  The cosmological
measurements will hence continue to 
benefit from higher halo survey density until $E\ll 0.05$.  Even with
optimal halo weighting this does not occur until 
$M_{\rm  min}<10^{12}M_\odot$.  In most of the relevant regime, the
optimal mass weighting require 10 or more times fewer surveyed
redshifts than the Poisson formulae would have suggested.

\begin{figure}
\resizebox{\hsize}{!}{
\includegraphics[angle=0]{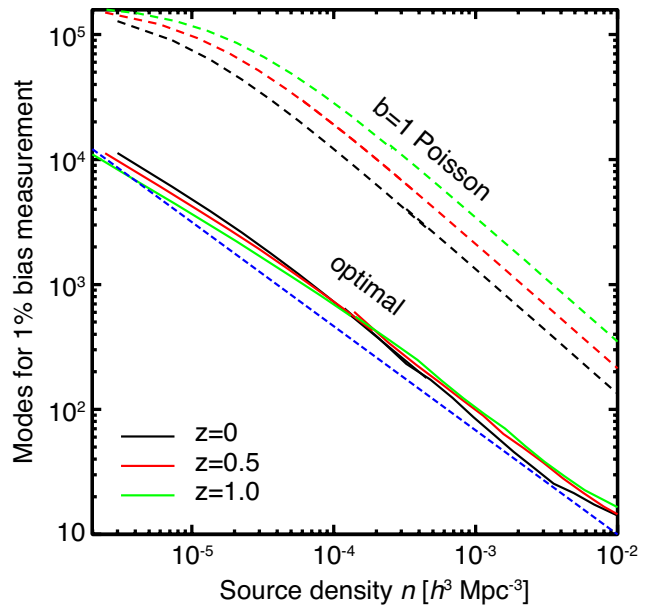}
}
\caption{The vertical axis is the number of modes $N_m$  that must be
  measured for a 1\% measure  of the bias of an optimal mass estimator using
  the cross-correlation of lensing and galaxy redshift survey.  It is
  plotted vs the density of sources in the redshift survey.  The solid
  lines are for an optimal survey which targets one galaxy in each
  halo  above some mass limit. The dashed lines assume that target
  galaxies are predominantly selected  from halos with
  $b\approx1$.  The dashed blue line follows $N_m\propto n^{-5/6}$,
  illustrating that the total number of galaxies $\propto nN_m$ required in
  the optimal survey is a weak function of survey depth.
}
\label{mvsn}
\end{figure}
 
Given the high source density required to saturate the accuracy in
$b$, we ask whether it is more efficient to conduct a deep survey or
a shallower survey covering more sky and hence more modes.
Figure~\ref{mvsn} plots the number of modes that must be observed to
determine $b$ to 1\% accuracy, vs the space density $n$ of halos
surveyed, as we lower $M_{\rm min}$ of the survey.   We find
$N_m\propto n^{-5/6}$ describes the results well.  The total
number of redshifts to be obtained in the survey is $nN_m\propto
n^{1/6}$, hence this measure of the survey's expense depends very
weakly on depth.  We also note that the source density required for
the bias measurement is nearly independent of redshift using the
optimal survey strategy.  In contrast, a survey of number-weighted
emission-line galaxies, plotted with the dashed lines, 
requires $>15\times$ higher $n$ than the optimal strategy at $z=0$,
and galaxy densities that increase to $z=1$.

\subsection{Galaxies as weights}
\label{LRG}

Most measurements of large-scale structure to date have used galaxy
densities as mass estimators.  A spectroscopic survey of galaxies can 
be thought of as weighting halos by the number of galaxies per halo 
\cite{Scoccimarro01}.  
The halo occupancy distribution (HOD) gives the probability $p(N|M)$ 
of finding $N$ galaxies in a halo of mass $M$.  

The use of galaxies as mass tracers will be inferior to an optimal
halo-weighting scheme, in the sense of having higher stochasticity $E$
for a given number of redshifts, for three reasons:
\begin{enumerate}
\item Ideally only one redshift per halo is needed, so a galaxy survey
  is in some sense wasting spectroscopic resources if more than one
  galaxy is targeted per halo.  
\item The mean HOD $g(M)\equiv \langle p(N|M) \rangle $ may
 not match the optimal halo weighting.  
\item The occupancy of a given halo is an integer drawn from the
HOD, which adds a source of stochasticity to the weight assignment 
 and can propagate into increased stochasticity in the mass estimation,
 even if the mean HOD is a close to the optimal weight. 
\end{enumerate}
The first penalty is typically not large: the HOD is typically divided
into the probability $f_c(M)$ of having a single central galaxy in the
halo, plus a distribution $p(N_s|M)$ of the number $N_s$ of satellite
galaxies.  The latter is typically taken to be a Poisson distribution
with a mean $\bar N_s(M)$.  For most galaxy samples, the fraction of
satellites is 10--20\%, a small perturbation to the number of redshift 
targets.

This also implies, however, that $\approx 90\%$ of the halos detected
in a survey are occupied by a single target galaxy, and hence given
equal weight.  We are then drawn to the second issue: does the mean
HOD $g(M)$ serve as a nearly-optimal weight function?  
In Figures~\ref{wvsm}, we plot the mean number of blue galaxies
(in cyan lines) and LRGs (in orange lines) in each halo as a function 
of halo mass, as determined from SDSS data.

For a simple HOD in which $f_c$ is a step function, the mean HOD is
\begin{equation}
 \label{wsdssxi}
 g(M) = 
\begin{cases}
 1 + \left(\frac{M}{M_1}\right)^{\alpha} & M\ge M_0 \\
 0 & M<M_0.
\end{cases}
\end{equation}
$M_1$, $\alpha$, and the cutoff $M_0$ are dependent upon the
luminosity cut or other criteria used to define the galaxy sample.
For luminosity-limited samples, typical values are
$M_1 \approx 20 M_0$ and $\alpha\approx 1$ \citep{Zehavi05, Zehavi10}. 

But note the similarity of this mean HOD functional form to
Equations~(\ref{woptsample}) and (\ref{halomodelw}) 
for the optimal weight.
The optimal weight for a catalog of halos with mass $M>M_{\rm min}$
is very well approximated by a function of the form
$w(M)= 1 + (M/\beta M_{\rm min})^{\alpha}$,
where $\alpha$ and $\beta$ depend on $M_{\rm min}$. 
For low $M_{\rm min}$, $\alpha \approx 1$, with $\alpha$ decreasing 
at higher $M_{\rm min}$.   For $M_{\rm min}$ in the range
$10^{11}$--$10^{13}M_\odot$ at $z=0$, values of $3<\beta<9$ yield 
the least stochasticity, with $E$ values indistinguishable from 
the optimal weighting.  

Hence by a useful coincidence, the mean HOD for luminosity-selected
galaxies bears close resemblance to an optimal halo weighting function, 
except that the HOD tends to have a longer flat low-mass plateau 
($\beta\approx 20$) than the optimal weight ($\beta\approx 4$).  
H10 noted that the optimal weight function they derived is 
well approximated by equation~(\ref{wsdssxi}), with $\beta\sim 3$, 
but they did not make the connection to galaxy HODs.  

\begin{figure}
 \resizebox{\hsize}{!}{
  \includegraphics[angle=0]{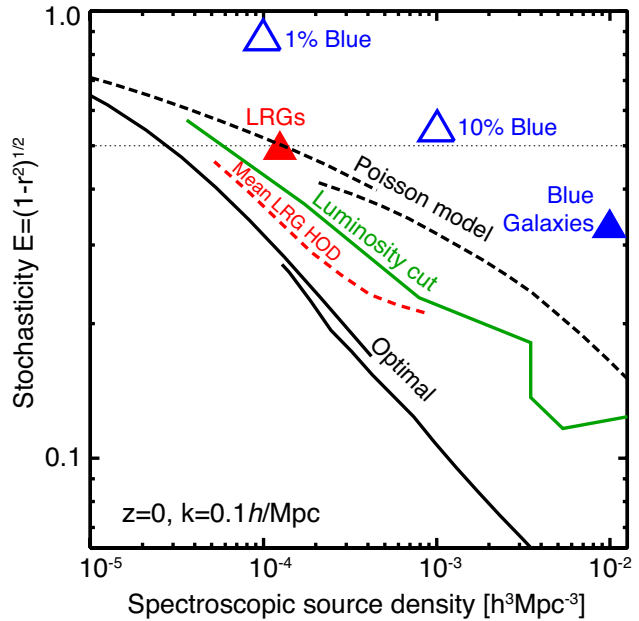}}
  \caption{Stochasticity $E$ as a function of the number density of
    redshifts obtained in the survey..  The black lines plot
           the result of an ideal mass-reconstruction strategy, in
           which one redshift is obtained for each halo above a cutoff
           mass, and an optimal $w(M)$ is applied to each surveyed
           halo.  The dotted black curve plots the stochasticity one
           would expect from this strategy if halos were a biased
           Poisson sampling of the mass---the optimal survey requires
           significantly fewer redshifts than predicted by the Poisson
           model for the same $E$.  The green line gives $E$
           resulting from weighting each halo by the number of
           galaxies above a luminosity cut---assuming galaxies occupy
           halos as per \citet{Zehavi10}.  The blue and red triangles
           result from galaxy-weighting using the HODs for blue and
           luminous red galaxies, respectively, in the SDSS.  The LRG
           and luminosity-cut surveys are worse than optimal primarily
           because of randomness in the halo occupancy; the dashed red
           line shows the result of eliminating this randomness by
           weighting each halo with the {\em mean} HOD.  Blue galaxies
           are very inefficient for reconstructing the mass; if
           emission-line spectroscopic galaxies are a random 10\% or
           1\% subsample of the blue population, they are also
           inefficient, not even attaining the $E=0.5$ required for a
           volume-limited power spectrum measurement.
}
\label{evsn}
\end{figure}

How far from optimal are the mean HODs?
We examine the case of luminous red galaxies (LRGs) first.
We obtain the mean number of LRGs from the HOD fitting results of 
\citet{Zheng09}, using equation~(B3) in their paper to model the 
dependence of model parameter on $\sigma_8$.  The mean HOD is 
shown in orange in Figure~(\ref{wvsm}).
Figure~\ref{evsn} plots the stochasticity $E$ vs the space density 
$n$ of target halos at $k=0.1h$Mpc$^{-1}$ and $z=0$, with the solid 
black line showing the best possible result from optimal targeting 
and weighting of halos ($M_{\rm min}$ is an implicit parameter for 
the black curve).  The dashed red line shows the result of using 
the mean LRG HOD as a halo weight.  A subtlety is that the LRG HOD 
does not have a step-function cutoff---the probability $f_c$ of a 
central galaxy follows an error function and hence has no single 
well-defined $M_{\rm min}$.  
The red dashed line shows the result of 
varying a low-mass cutoff applied to the LRG HOD.
We find that the $E$ vs $n$ behavior is within 10--20\% of the optimal 
result as long as we cut out halos with $f_c<0.1$.  
For LRGs, at least, the mean HOD is therefore a good
choice of weight function.  We will examine other galaxy
classes below.

To examine the impact of item (iii), the stochasticity of the HOD, on
mass-estimator performance, we populate the halos in the Millennium 
simulation with LRGs as per the HOD prescription.  We place central 
galaxies in the specified fraction of halos, plus a Poisson-distributed 
number of satellite galaxies in the halos which host a central galaxy.  
The red triangle in Figure~\ref{evsn} shows the stochasticity $E$ vs 
the number density of LRGs.  We find stochastically occupied halos
achieve a factor of two higher (worse) $E$ than the deterministic
weighting by the mean HOD.  While the stochastic LRG HOD requires
$\approx 3\times$ as many redshifts to reach $E=0.5$ as an ideal
survey would require, it is similar to what one would predict from an
optimal survey if the biased-Poisson model were a correct description
of halo stochasticity (dashed black line).

The green solid curve in Figure~\ref{evsn} shows the result of
weighting the halos by the number of galaxies drawn from HODs with
varying minimum galaxy luminosities, from \citet{Zehavi10}.  The
behavior is similar to the LRG HOD: even though the mean number of 
galaxies per halo looks like the optimal weight, using the actual 
number of galaxies as weight results in larger $E$.  
The randomness in the number of galaxies in each halo introduces 
additional stochasticity that degrades the mass estimator.  

If galaxies are to be used to provide optimal reconstructions of 
the mass, then they must themselves be weighted in some way so as 
to reduce the stochasticity in the weight applied to halos of a 
given mass.  Determining the optimal mark is an interesting problem 
for the future. Formalism for treating this more general problem 
has been developed in \citet{Sheth05}, and can be used directly, 
but is beyond the scope of this work.  

Our results suggest that it is interesting to study how best to supplement 
the spectroscopic galaxies with a larger, deeper photometric-redshift 
sample.  The spectroscopic galaxies can be weighted by the number of 
photo-z galaxies consistent with sharing the same halo.  The deeper 
photo-z catalog potentially has lower stochasticity in halo mass 
estimates.

\begin{figure}
\resizebox{\hsize}{!}{
\includegraphics[angle=0]{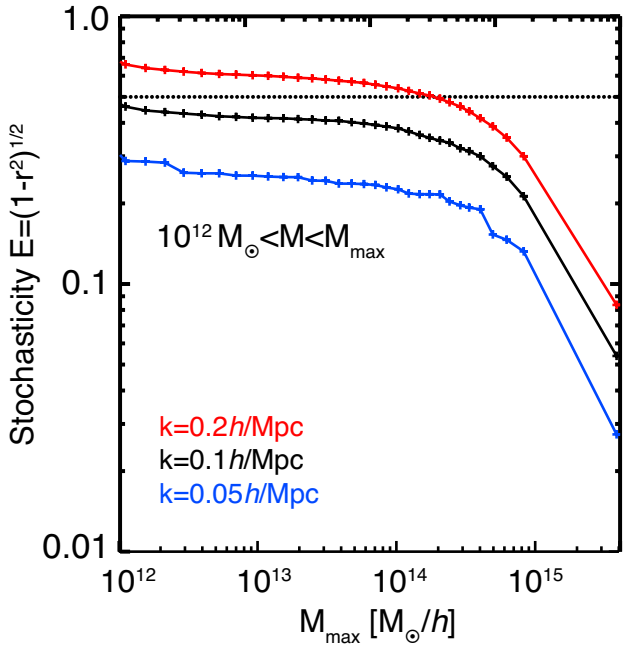}
}
\caption{We plot the stochasticity of an optimal mass estimator
  created from a catalog of halos in the Millennium simulation in the
  mass range 
$10^{12} h^{-1}  M_\odot < M < M_{\rm max}$ vs the {\em
    upper} limit $M_{\rm max}$  of the halo catalog. This
  demonstrates that one cannot improve the stochasticity of the
  estimator below some floor unless one detects (and heavily weights)
  the rare halos above $\approx 10^{14}h^{-1} M_\odot$.
}
\label{rollup}
\end{figure}
 
\subsubsection{Surveys with blue or emission-line galaxies}
We have seen that the mean HODs for LRGs and luminosity-selected
galaxies are good approximations to the optimal weight, but the
stochasticity in halo occupation degrades $E$ for a given source
density $n$.  Galaxy redshift surveys based on emission line detection
will likely result in substantially different halo weightings, so we
investigate the mass-reconstruction performance of such a survey
relative to an LRG survey or optimal weighting.

We model the emission-line sample by starting with the mean HOD for
blue galaxies
given by equations~(10) and~(11) and Table~4 in \citep{Zehavi05}:
\begin{equation}
 g_{blue}(m) = \left(\frac{m}{M_{B}}\right)^{0.8} 
              + 0.7\,{\rm e}^{-[2\log(M/10^{12} h^{-1}M_\odot)]^2}
\end{equation}
where $M_B = 7\times10^{13} h^{-1}M_{\odot}$ and $\alpha_B=0.8$ 
\citep[following][]{Sheth01}.  This is plotted as the cyan line in
Figure~\ref{wvsm}.
There is a bump in the 
number of blue galaxies between $\sim M^{11}~h^{-1}M_{\odot}$ to 
$\sim M^{12}~h^{-1}M_{\odot}$, which is very different from 
the optimal weight. The outcome of $E$ from weighting halos in the
Millennium simulation according to galaxy counts drawn from this HOD 
is shown in the 
blue triangle of Figure~(\ref{evsn}). Although the blue galaxy sample 
achieves lower $E$ than the LRG does, notice that it requires $100\times$ 
more redshifts, i.e., is $100\times$ more costly than an optimally weighted 
sample.  If we were to under-sample the blue galaxies---e.g. by
obtaining redshifts for ten  percent, or one percent of the sample
with the brightest emission lines---then $E$ would rise to 0.54 and 
0.86, respectively, if the sub-sampling rate is independent of halo
properties. Clearly, emission line samples are a very inefficient 
way to reconstruct the mass, although this disadvantage is countered
by the fact that emission lines can be much stronger and easily
detected relative to LRG absorption features.  Optimization of a
survey would need to weigh these effects.

\subsubsection{Surveys that under-weight massive clusters}
Galaxies in massive clusters tend to be strongly deficient in 21-cm
emission and other gas-phase emission lines.  A redshift survey
selected by such criteria will tend to miss or under-weight the most
massive clusters. In the Figure~\ref{rollup}, we show that the 
absence of high-mass halos from a catalog sets a floor on the 
attainable stochasticity even with optimal mass estimation. When 
halos with $\sim 10^{12}h^{-1}M_{\odot}<M<M_{\rm max}$ are detected by 
the survey, we find that $E \approx 0.5$ is achievable at 
$z=0$, $k=0.2h$Mpc$^{-1}$ without detecting massive halos.
However, cosmological inferences requiring lower values of $E$ 
require a means of detecting and appropriately weighting clusters
above $10^{14}h^{-1}M_\odot$.  Surveys without the ability to identify
massive clusters reach a limit in $E$ that cannot be improved by
addition of more low-mass halo detections.  

The floor on $E$ can be estimated using equation~(\ref{E2sample}).
Consider the case when our halo catalog only contains objects 
{\em below} some $m_d$, so the ``dust'' is the mass in massive objects.  
Then $\sum_j n_jv_j^2$ can be large, so $E_{\rm pois}\to 0$.  
$\noise_m$ is dominated by the most massive objects: at $z=0$, about 
90\% of $\noise_m$ is contributed by objects with masses above 
$10^{14}h^{-1}M_\odot$ (we have assumed $\sigma_8 = 0.9$).  
If these are missing from the catalog, then $\noise_d\to \noise_m$, 
making $E_{\rm opt}\approx (1 + P/\noise_m)^{-1/2}$; as a result, 
on scales of $k\sim 0.1h$Mpc$^{-1}$, $E_{\rm opt}$ cannot be made 
smaller than about 0.25.

\section{Conclusion and Discussion}\label{discuss}
We have determined the weighting scheme that minimizes the 
stochasticity between the linearly weighted halo field and the 
associated mass density field.  The optimal weight function depends 
on the mass range of the halo catalog, how much mass is missing from 
the halo catalogue, and how the halos cluster.  (I.e., we showed how 
the weight is modified by a halo-mass dependent selection function.)
We show that neither mass weighting nor bias weighting of the halos 
is optimal. The first principal component of the halo covariance
matrix $\mC$ is also 
usually not the optimal, although H10 show that the weakest PC of the
altered noise matrix $\mC-\vb P \vb^T$ is a good approximation to an optimal
weight 
when the numbers of halos in all bins are equal.
Rather, we demonstrate that, under very general circumstances, 
the optimal weight will be a mix of bias weighting and mass weighting, 
simply because the mass is comprised of the mass-weighted halo catalog 
plus the mass in the ``dust bin'' of structures below the halo detection
threshold.  

The halo model can generally give a reasonably good description for 
the optimal weight function and its associated stochasticity with 
two important alterations: first, it is necessary to treat the 
halos as though they sample a continuous ``halo field'' that is 
distinct from the mass distribution, and that they have a bias 
$v(M)$ with respect to the halo field that differs from the bias 
$b(M)$ with respect to mass.  
Second, halo catalogs extending below $10^{12}h^{-1}M_{\odot}$ do not 
reconstruct the mass as well as the halo model predicts.
However, we find that the model generally overestimates the optimal 
stochasticity, even on the large scales where one might have expected 
to find good agreement.  We suspect this is due to the combined effects 
of non-linear bias and halo exclusion, which our treatment currently 
ignores.  

We also note that the randomness in the halo shapes at fixed mass---i.e. 
ellipticity, concentration, and/or substructure---introduces stochasticity 
into a mass estimator built from a halo catalog in which the halo profiles 
are reduced to points, setting a lower limit on the attainable $E$.  
In the Millennium catalog, this structure stochasticity substantially
degrades the mass estimation for $k>0.2h\,{\rm Mpc}^{-1}$.
Since information about halo shapes is difficult to obtain in 
observations, the lower limit on $E$ from point-like halos in simulations 
is also the best one can achieve in real observations -- although, 
because galaxies are expected to be reasonably faithful tracers of 
halo profiles, it may be that they can be used to further reduce $E$
into the non-linear regime.  

An optimally weighted halo catalog can have an effective number
density $(nb^2)_{\rm eff}$ up to $15\times$ better (higher) than one
would have predicted for the same halo catalog in a biased-Poisson
model of halo stochasticity.  This gain means that a volume-limited
measurement of the linear-regime power spectrum of matter for the
entire observable $z<1$ universe could in principal be accomplished
with only 6 million spectroscopic redshift measurements.  Such a
program would require outside information, perhaps a deep imaging
photo-z survey, to identify halos and provide reliable mass estimates
or marks to apply to spectroscopic targets.
(See H10 for an estimate of the effect of mass-estimator 
degradation due to a generic log-normal error distribution in the 
estimation of halo masses.)

We use halo occupancy distribution models to estimate the 
stochasticity $E$ resulting from more traditional surveys which 
apply uniform weights to targeted galaxies.  
The mean HODs for luminosity-thresholded samples and LRGs are
remarkably useful approximations to 
the optimal weight functions.  However, additional 
stochasticity is introduced into the mass estimator by the random 
variations in halo occupancy about the mean.  
Hence luminosity-selected or LRG catalogs require $\approx 3\times$
more redshifts to reach a given $E$ than a survey with perfect
knowledge of halo masses for which the optimal weighting can be 
applied.  In contrast, HODs for blue or emission-line galaxies do 
not resemble the optimal weights, and hence require 
$\approx 100\times$ more redshift measurements than an 
optimally-weighted survey to obtain a given $E$.  Random sub-sampling 
such galaxies to the same space density as LRGs yields $E$ values 
that are $2\times$ larger than those for LRGs.  
We also find that low-mass halos cannot reconstruct the shot noise 
contributed by the massive halos, setting an upper limit on the 
fidelity of the mass reconstruction for surveys that fail to identify 
the most massive clusters.

Application of optimal halo weighting can be even more beneficial 
for studies of cross-correlation between gravitational potential
(i.e. mass) and other cosmological signals, since these experiments
gain rapidly as the stochasticity $E$ drops below the $E=0.5$ needed
to make volume-limited power-spectrum measurements.  
\citet{Park10}, find, for example, that mass weighting halos can greatly  
accuracy in estimation of gravitational potential if the halo catalog
extends down to $\approx10^{13}h^{-1}M_\odot$ or lower.  

Applying the optimal weight is obviously efficient in reducing noise in 
the estimation of BAO from power spectra and in cross-correlation
cosmological tests.  In future work we will extend this study to the 
use of redshift space distortions to measure the growth rate of 
structure \citep[e.g.][]{Okumura10}.  We are also investigating the
potential for galaxy marking and non-linear mass estimators to further
improve the ability to trace large-scale structure with observational
data.

\section*{ACKNOWLEDGMENT}
We thank Roman Scoccimarro for providing the NYU simulations. This work is supported by DOE grant DE-FG02-95ER40893, and grants AST-0908027 and AST-0908241 from the National Science Foundation. The Millennium simulation used in this paper was carried out as part of the programme of the Virgo Consortium on the Regatta supercomputer of the Computing Centre of the Max-Planck-Society in Garching. We thanks John Helly for helping accessing the Millennium database. YC thanks the hospitality of the Institute for Computational Cosmology in Durham University when this work was finishing. 

\bibliographystyle{mn2e}
\bibliography{HaloBias}

\begin{thebibliography}{}

\bibitem[\protect\citeauthoryear{{Abbas} \& {Sheth}}{{Abbas} \&
  {Sheth}}{2007}]{Abbas07}
{Abbas} U.,  {Sheth} R.~K.,  2007, \mnras, 378, 641

\bibitem[\protect\citeauthoryear{{Bardeen}, {Bond}, {Kaiser} \&
  {Szalay}}{{Bardeen} et~al.}{1986}]{Bardeen86}
{Bardeen} J.~M.,  {Bond} J.~R.,  {Kaiser} N.,    {Szalay} A.~S.,  1986, \apj,
  304, 15

\bibitem[\protect\citeauthoryear{{Bernstein} \& {Jain}}{{Bernstein} \&
  {Jain}}{2004}]{Bernstein04}
{Bernstein} G.,  {Jain} B.,  2004, \apj, 600, 17

\bibitem[\protect\citeauthoryear{{Bonoli} \& {Pen}}{{Bonoli} \&
  {Pen}}{2009}]{Bonoli09}
{Bonoli} S.,  {Pen} U.~L.,  2009, \mnras, 396, 1610

\bibitem[\protect\citeauthoryear{{Cole} \& {Kaiser}}{{Cole} \&
  {Kaiser}}{1989}]{Cole89}
{Cole} S.,  {Kaiser} N.,  1989, \mnras, 237, 1127

\bibitem[\protect\citeauthoryear{{Davis}, {Efstathiou}, {Frenk} \&
  {White}}{{Davis} et~al.}{1985}]{Davis85}
{Davis} M.,  {Efstathiou} G.,  {Frenk} C.~S.,    {White} S.~D.~M.,  1985, \apj,
  292, 371

\bibitem[\protect\citeauthoryear{{Dekel} \& {Lahav}}{{Dekel} \&
  {Lahav}}{1999}]{DL99}
{Dekel} A.,  {Lahav} O.,  1999, \apj, 520, 24

\bibitem[\protect\citeauthoryear{{Gladders} \& {Yee}}{{Gladders} \&
  {Yee}}{2000}]{RCS}
{Gladders} M.~D.,  {Yee} H.~K.~C.,  2000, \aj, 120, 2148

\bibitem[\protect\citeauthoryear{{Hamaus}, {Seljak}, {Desjacques}, {Smith} \&
  {Baldauf}}{{Hamaus} et~al.}{2010}]{Hamaus10}
{Hamaus} N.,  {Seljak} U.,  {Desjacques} V.,  {Smith} R.~E.,    {Baldauf} T.,
  2010, ArXiv e-prints

\bibitem[\protect\citeauthoryear{{Hockney} \& {Eastwood}}{{Hockney} \&
  {Eastwood}}{1981}]{Hockney81}
{Hockney} R.~W.,  {Eastwood} J.~W.,  1981, {Computer Simulation Using
  Particles}

\bibitem[\protect\citeauthoryear{{Koester}, {McKay}, {Annis}, {Wechsler},
  {Evrard}, {Rozo}, {Bleem}, {Sheldon} \& {Johnston}}{{Koester}
  et~al.}{2007}]{MaxBCG}
{Koester} B.~P.,  {McKay} T.~A.,  {Annis} J.,  {Wechsler} R.~H.,  {Evrard}
  A.~E.,  {Rozo} E.,  {Bleem} L.,  {Sheldon} E.~S.,    {Johnston} D.,  2007,
  \apj, 660, 221

\bibitem[\protect\citeauthoryear{{Manera}, {Sheth} \& {Scoccimarro}}{{Manera}
  et~al.}{2010}]{Manera10}
{Manera} M.,  {Sheth} R.~K.,    {Scoccimarro} R.,  2010, \mnras, 402, 589

\bibitem[\protect\citeauthoryear{{Mo} \& {White}}{{Mo} \& {White}}{1996}]{MW96}
{Mo} H.~J.,  {White} S.~D.~M.,  1996, \mnras, 282, 347

\bibitem[\protect\citeauthoryear{{Okumura} \& {Jing}}{{Okumura} \&
  {Jing}}{2010}]{Okumura10}
{Okumura} T.,  {Jing} Y.~P.,  2010, ArXiv e-prints

\bibitem[\protect\citeauthoryear{{Park} \& {Choi}}{{Park} \&
  {Choi}}{2009}]{Park09}
{Park} C.,  {Choi} Y.,  2009, \apj, 691, 1828

\bibitem[\protect\citeauthoryear{{Park}, {Kim} \& {Park}}{{Park}
  et~al.}{2010}]{Park10}
{Park} H.,  {Kim} J.,    {Park} C.,  2010, \apj, 714, 207

\bibitem[\protect\citeauthoryear{{Pen}}{{Pen}}{2004}]{Pen04}
{Pen} U.,  2004, \mnras, 350, 1445

\bibitem[\protect\citeauthoryear{{Percival}, {Verde} \& {Peacock}}{{Percival}
  et~al.}{2004}]{Percival04}
{Percival} W.~J.,  {Verde} L.,    {Peacock} J.~A.,  2004, \mnras, 347, 645

\bibitem[\protect\citeauthoryear{{Scoccimarro}}{{Scoccimarro}}{1998}]{Scoccima%
rro98}
{Scoccimarro} R.,  1998, \mnras, 299, 1097

\bibitem[\protect\citeauthoryear{{Scoccimarro}, {Sheth}, {Hui} \&
  {Jain}}{{Scoccimarro} et~al.}{2001}]{Scoccimarro01}
{Scoccimarro} R.,  {Sheth} R.~K.,  {Hui} L.,    {Jain} B.,  2001, \apj, 546, 20

\bibitem[\protect\citeauthoryear{{Seljak}, {Hamaus} \& {Desjacques}}{{Seljak}
  et~al.}{2009}]{SHD09}
{Seljak} U.,  {Hamaus} N.,    {Desjacques} V.,  2009, Physical Review Letters,
  103, 091303

\bibitem[\protect\citeauthoryear{{Seljak} \& {Warren}}{{Seljak} \&
  {Warren}}{2004}]{Seljak04}
{Seljak} U.,  {Warren} M.~S.,  2004, \mnras, 355, 129

\bibitem[\protect\citeauthoryear{{Seljak} \& {Zaldarriaga}}{{Seljak} \&
  {Zaldarriaga}}{1996}]{Seljak96}
{Seljak} U.,  {Zaldarriaga} M.,  1996, \apj, 469, 437

\bibitem[\protect\citeauthoryear{{Sheth}}{{Sheth}}{2005}]{Sheth05}
{Sheth} R.~K.,  2005, \mnras, 364, 796

\bibitem[\protect\citeauthoryear{{Sheth}, {Connolly} \& {Skibba}}{{Sheth}
  et~al.}{2005}]{SCS05}
{Sheth} R.~K.,  {Connolly} A.~J.,    {Skibba} R.,  2005, ArXiv Astrophysics
  e-prints

\bibitem[\protect\citeauthoryear{{Sheth} \& {Diaferio}}{{Sheth} \&
  {Diaferio}}{2001}]{Sheth01}
{Sheth} R.~K.,  {Diaferio} A.,  2001, \mnras, 322, 901

\bibitem[\protect\citeauthoryear{{Sheth} \& {Jain}}{{Sheth} \&
  {Jain}}{2003}]{ShethJain03}
{Sheth} R.~K.,  {Jain} B.,  2003, \mnras, 345, 529

\bibitem[\protect\citeauthoryear{{Sheth} \& {Lemson}}{{Sheth} \&
  {Lemson}}{1999}]{Sheth99b}
{Sheth} R.~K.,  {Lemson} G.,  1999, \mnras, 304, 767

\bibitem[\protect\citeauthoryear{{Sheth} \& {Tormen}}{{Sheth} \&
  {Tormen}}{1999}]{Sheth99}
{Sheth} R.~K.,  {Tormen} G.,  1999, \mnras, 308, 119

\bibitem[\protect\citeauthoryear{{Sheth} \& {Tormen}}{{Sheth} \&
  {Tormen}}{2002}]{ST02}
{Sheth} R.~K.,  {Tormen} G.,  2002, \mnras, 329, 61

\bibitem[\protect\citeauthoryear{{Smith}, {Scoccimarro} \& {Sheth}}{{Smith}
  et~al.}{2007}]{Smith07}
{Smith} R.~E.,  {Scoccimarro} R.,    {Sheth} R.~K.,  2007, \prd, 75, 063512

\bibitem[\protect\citeauthoryear{{Springel}}{{Springel}}{2005}]{Springel05a}
{Springel} V.,  2005, \mnras, 364, 1105

\bibitem[\protect\citeauthoryear{{Springel}, {White}, {Jenkins}, {Frenk},
  {Yoshida}, {Gao}, {Navarro}, {Thacker}, {Croton}, {Helly}, {Peacock}, {Cole},
  {Thomas}, {Couchman}, {Evrard}, {Colberg} \& {Pearce}}{{Springel}
  et~al.}{2005}]{Springel05b}
{Springel} V.,  {White} S.~D.~M.,  {Jenkins} A.,  {Frenk} C.~S.,  {Yoshida} N.,
   {Gao} L.,  {Navarro} J.,  {Thacker} R.,  {Croton} D.,  {Helly} J.,
  {Peacock} J.~A.,  {Cole} S.,  {Thomas} P.,  {Couchman} H.,  {Evrard} A.,
  {Colberg} J.,    {Pearce} F.,  2005, \nat, 435, 629

\bibitem[\protect\citeauthoryear{{Sunyaev} \& {Zeldovich}}{{Sunyaev} \&
  {Zeldovich}}{1972}]{Sunyaev72}
{Sunyaev} R.~A.,  {Zeldovich} Y.~B.,  1972, Comments on Astrophysics and Space
  Physics, 4, 173

\bibitem[\protect\citeauthoryear{{Tegmark}, {Taylor} \& {Heavens}}{{Tegmark}
  et~al.}{1997}]{TTH}
{Tegmark} M.,  {Taylor} A.~N.,    {Heavens} A.~F.,  1997, \apj, 480, 22

\bibitem[\protect\citeauthoryear{{White}}{{White}}{1996}]{white96}
{White} S.~D.~M.,  1996, {in Schaeffer R., Silk J., Spiro M., Zinn-Justin J.,
  eds, Cosmology and Large-Scale Structure, Elsevier, Amsterdam p. 349}

\bibitem[\protect\citeauthoryear{{Zehavi}, {Zheng}, {Weinberg}, {Blanton},
  {Bahcall}, {Berlind}, {Brinkmann} \& {Frieman}}{{Zehavi}
  et~al.}{2010}]{Zehavi10}
{Zehavi} I.,  {Zheng} Z.,  {Weinberg} D.~H.,  {Blanton} M.~R.,  {Bahcall}
  N.~A.,  {Berlind} A.~A.,  {Brinkmann} J.,    {Frieman} J.~A. e.~a.,  2010,
  ArXiv e-prints

\bibitem[\protect\citeauthoryear{{Zehavi}, {Zheng}, {Weinberg}, {Frieman},
  {Berlind}, {Blanton}, {Scoccimarro} \& {Sheth}}{{Zehavi}
  et~al.}{2005}]{Zehavi05}
{Zehavi} I.,  {Zheng} Z.,  {Weinberg} D.~H.,  {Frieman} J.~A.,  {Berlind}
  A.~A.,  {Blanton} M.~R.,  {Scoccimarro} R.,    {Sheth} R.~K. e.~a.,  2005,
  \apj, 630, 1

\bibitem[\protect\citeauthoryear{{Zheng}, {Zehavi}, {Eisenstein}, {Weinberg} \&
  {Jing}}{{Zheng} et~al.}{2009}]{Zheng09}
{Zheng} Z.,  {Zehavi} I.,  {Eisenstein} D.~J.,  {Weinberg} D.~H.,    {Jing}
  Y.~P.,  2009, \apj, 707, 554

\end{thebibliography}

\end{document}